\documentclass[aps,prev,twocolumn,preprintnumbers,floatfix,nofootinbib]{revtex4-1}

\usepackage{graphicx}
\usepackage[ngerman,english]{babel}
\usepackage{bm}
\usepackage{times}
\usepackage{slashed}
\usepackage{mathtools}
\usepackage{color}
\usepackage{epsfig}
\usepackage{dcolumn}
\usepackage{textcomp}
\usepackage{color}
\newcommand\scalemath[2]{\scalebox{#1}{\mbox{\ensuremath{\displaystyle #2}}}}

\begin{document} 

\title{Collider Bounds on 2-Higgs Doublet Models with $U(1)_X$ Gauge Symmetries}

\author{Daniel A. Camargo$^1$}\email{dacamargov@gmail.com}
\author{Luigi Delle Rose$^2$}\email{luigi.dellerose@fi.infn.it}
\author{Stefano Moretti$^3$}\email{s.moretti@soton.ac.uk}
\author{Farinaldo S. Queiroz$^{1}$}\email{farinaldo.queiroz@iip.ufrn.br}

\affiliation{$^1$International Institute of Physics, Universidade Federal do Rio Grande do Norte, Campus Universitario, Lagoa Nova, Natal-RN 59078-970, Brazil}
\affiliation{$^2$INFN, Sezione di Firenze, and Dipartimento di Fisica ed Astronomia, Universit\a di Firenze,  Via G. Sansone 1, 50019 Sesto Fiorentino, Italy}
\affiliation{$^3$School of Physics and Astronomy, University of Southampton, Highfield, Southampton SO17 1BJ,
United Kingdom}

\begin{abstract}
\noindent
2-Higgs Doublet Models (2HDMs) typically need to invoke an ad-hoc discrete symmetry to avoid severe flavor bounds and in addition feature massless neutrinos, thus falling short of naturally complying with existing data. However, when augmented by an Abelian gauge symmetry naturally incorporating neutrino masses via a type-I seesaw mechanism while at the same time escaping flavor changing interactions, such enlarged 2HDMs become very attractive phenomenologically. In such frameworks, the distinctive element is the $Z'$ gauge boson generated by the spontaneous breaking of the Abelian group $U(1)_X$. In this work, we derive updated collider bounds on it.   
Several theoretical setups are possible, each with different and sometimes suppressed couplings to quarks and  leptons.  Thus, complementary data from dijet and dilepton resonance searches need to be considered to fully probe these objects. We employ the corresponding datasets as obtained at the Large Hadron Collider (LHC) at the 13 TeV CMs energy for $\mathcal{L}=12,36$ and $300$ fb$^{-1}$ of luminosity.   Moreover, we present the potential sensitivity to such $Z'$s of the High Luminosity LHC (HL-LHC) and High Energy LHC (HE-LHC).
\end{abstract}

\maketitle
\flushbottom

\section{Introduction}
\label{sec_introduction}

The Standard Model (SM) offers the best description of strong and Electro-Weak (EW) interactions  and has successfully passed all precision tests up to now \cite{Novaes:1999yn}. In the SM, fermion masses are generated via the presence of one scalar doublet which gives rise to the $125$~GeV Higgs boson discovered some years ago at the LHC \cite{Chatrchyan:2012xdj,Aad:2012tfa}. One of the precision measurements that attest the predictability of the SM concerns the $\rho$ parameter, 
$\rho= M_W^2/(M_Z^2 \cos \theta_W^2)$, where $\theta_W$ is the Weinberg angle. 
Today, the EW tests lead to $\rho=1.01032\pm 0.00009$ \cite{Patrignani:2016xqp}, with the error bar being driven mostly by the uncertainty of the top quark mass which appears at one loop level in the calculations.\\

In general extended scalar sectors the $\rho$ parameter is given by 
\begin{equation}
\rho = \frac{{\displaystyle \sum_{i=1}^n} \left[
I_i \left( I_i+1 \right) - \frac{1}{4}\, Y_i^2 \right] v_i}
{{\displaystyle \sum_{i=1}^n}\, \frac{1}{2}\, Y_i^2 v_i}, 
\label{jduei}
\end{equation}
where $I_i$ and $Y_i$ are the isospin and hypercharge of a scalar representation with Vacuum Expectation Value (VEV) $v_i$. 
Hence models with extended Higgs sectors that feature scalar doublets with hypercharge equal to unity or scalar singlets with zero hypercharge straightforwardly preserve $\rho=1$ at tree level. From this perspective, models with the presence of a second Higgs doublet stand out because they also offer a hospitable environment for EW phase transition \cite{Davies:1994id,Cline:1995dg,Bai:2012ex,Barger:2013ofa,Dumont:2014wha}, collider \cite{Alves:2016bib} and flavor physics \cite{Atwood:1996vj,Crivellin:2015mga,Lindner:2016bgg} (see \cite{Branco:2011iw} for a review). \\

However, such 2HDMs \cite{Lee:1973iz} suffer from severe flavor bounds because the presence of a second Higgs doublet induces flavor changing neutral currents (FCNCs) \cite{Paschos:1976ay,Glashow:1976nt}. One way around this problem has been to work with Yukawa couplings that obey certain relations so that FCNCs are suppressed \cite{Ferreira:2010ir,Felipe:2014zka,Alves:2018kjr}. An orthogonal solution to this problem has been to enforce an ad-hoc $Z_2$ symmetry under which one scalar doublet is even and the other is odd.  That said, it would be elegant if one could realize this discrete symmetry via gauge principles and, in addition, also accommodate  neutrino masses since their explanation is still absent in the original formulation of the 2HDM \cite{Atwood:2005bf,Davidson:2010sf,Chao:2012pt,Kanemura:2013qva,Liu:2016mpf,Bertuzzo:2017sbj}.\\

Having that in mind, the relation between continuous symmetries and discrete symmetries has been addressed in \cite{Ferreira:2010ir,Serodio:2013gka}. Gauge symmetries have been considered in several 2HDM scenarios in \cite{Ko:2013zsa,Ko:2014uka,Berlin:2014cfa,Huang:2015wts,DelleRose:2017xil}, but the aforementioned flavor problem was addressed in this context only in \cite{Ko:2012hd} and was later expanded in \cite{Campos:2017dgc} with additional $U(1)_X$ models. In the presence of an additional $U(1)_X$ group, 
neutrino masses are naturally taken into account since the extra gauge symmetry requires three right-handed neutrinos in order to cancel the gauge anomalies \cite{Ko:2012hd,Campos:2017dgc}. Indeed, the corresponding Majorana mass term leads, through a type-I seesaw mechanism, to a simple explanation of the neutrino mass problem. 
2HDMs with a $U(1)_X$ gauge symmetry are also characterized by a richer phenomenology 
due to the presence of a massive $Z^\prime$ gauge boson that arises after the spontaneous symmetry breaking. Moreover, neutrino masses and dark matter have been addressed recently with gauge symmetries and a singlet scalar extension in \cite{Bauer:2018egk}. \\ 
Our goal in this work is to use dijet \cite{Sirunyan:2016iap} and dilepton \cite{Chatrchyan:2012xdj,Khachatryan:2016zqb,Aaboud:2016cth,Aaboud:2017buh} data from the LHC to constrain the mass of such $Z^\prime$ boson and, consequently, the viable parameter space of the model. Instead of focusing on only one $U(1)_X$ realization, we will investigate all eight $U(1)_X$ models introduced in \cite{Campos:2017dgc} which are capable of solving the flavor problem in the 2HDM as well as generating neutrino masses from gauge principles.\\

Our work is structured as follows. In {Section} \ref{sec_model} we review the models, in {Section} \ref{sec_collider} we derive the aforementioned collider limits, in {Section} \ref{sec_HL-HE-LHC} we comment on the sensitivity at the HL/HE-LHC. We draw our conclusions in {Section} \ref{sec_conclusion}.

\begin{table*}[!t]
\centering
\begin{tabular}{ccccccccc}
\hline 
Fields & $u_R$ & $d_R$ & $Q_L$ & $L_L$ & $e_R$ & $N_R$ & $\Phi _2$  & $\Phi_1$ \\ \hline 
Charges & $u$ & $d$ & $\frac{(u+d)}{2}$ & $\frac{-3(u+d)}{2}$ & $-(2u+d)$ & $-(u+2d)$ & $\frac{(u-d)}{2}$ & $\frac{5u}{2} +\frac{7d}{2}$  \\
$U(1)_{A}$ & $1$ & $-1$ & $0$ & $0$ & $-1$ & $1$ & $1$ & $-1$\\
$U(1)_{B}$ & $-1$ & $1$ & $0$ & $0$ & $1$ & $-1$ & $-1$ & $1$\\
$U(1)_{C}$ & $1/2$ & $-1$ & $-1/4$ & $3/4$ & $0$ & $3/2$ & $3/4$ & $9/4$\\
$U(1)_{D}$ & $1$ & $0$ & $1/2$ & $-3/2$ & $-2$ & $-1$ & $1/2$ & $5/2$\\ 
$U(1)_{E}$ & $0$ & $1$ & $1/2$ & $-3/2$ & $-1$ & $-2$ & $7/2$ & $-1/2$\\
$U(1)_{F}$ & $4/3$ & $2/3$ & $1$ & $-3$ & $-4$ & $-8/3$ & $1/3$ & $17/3$\\
$U(1)_{G}$ & $-1/3$ & $2/3$ & $1/6$ & $-1/2$ & $0$ & $-1$ & $-1/2$ & $-3/2$\\
$U(1)_{B-L}$ & $1/3$ & $1/3$ & $1/3$ & $-1$ & $-1$ & $-1$ & $0$ & $2$\\
\hline
\end{tabular}
\caption{Models capable of explaining neutrino masses and the absence of FCNCs in the $U(1)$ extended type I 2HDM.}
\label{cargas_u1_2hdm_tipoI}
\end{table*}

\section{The Model}
\label{sec_model}

The 2HDM is characterized by a rich collider and flavor phenomenology 
but suffers, in general, from FCNC bounds and lacks  an explanation for neutrino masses \cite{Lee:1973iz}. The 2HDM embedding Abelian gauge groups that solve these problems appeared in \cite{Ko:2012hd,DelleRose:2017xil,Campos:2017dgc}. There are several possible $U(1)_X$ symmetries that can be incorporated in the 2HDM that suppress FCNCs (see table \ref{cargas_u1_2hdm_tipoI}) while being free from triangle anomalies
through the addition of three right-handed neutrinos.
If the $U(1)_X$ charges of the two Higgs doublets are different, the Abelian gauge symmetry naturally prohibits one of the two doublets to participate in the generation of the SM fermion masses. This mechanism replaces the ad-hoc $Z_2$ discrete symmetry, which is commonly employed in the 2HDM, and provides a Yukawa structure similar to the type I scenario. 
Furthermore, one can properly choose the $U(1)_X$ quantum numbers of all particles such that no anomalies are present. In this respect, three right-handed neutrinos are necessary, which charges are fixed to $-(u+2 d)$, where $u$ and $d$ are the quantum numbers of the right-handed up- and down-quarks, respectively. For instance, if $u = d = \frac 13$, we end up with the well-known $B-L$ gauge symmetry. However, as seen in table \ref{cargas_u1_2hdm_tipoI}, many other models are also possible. \\
The SM fermions and the neutrinos acquire mass through the Lagrangian
\begin{equation}
\begin{split}
\label{2hdm_tipoI_u1}
\mathcal{L} _{Y _{\text{2HDM}}} &= y_2 ^d \bar{Q} _L \Phi _2 d_R + y_2 ^u \bar{Q} _L \widetilde \Phi _2 u_R + y_2 ^e \bar{L} _L \Phi _2 e_R \\
&+ y^{D} \bar{L} _{L}\widetilde \Phi _2 N_{R}+ Y^{M}\overline{(N_{R})^{c}}\Phi_{s}N_{R}+
h.c. \\
\end{split}
\end{equation}
Notice that only the scalar doublet $\Phi _2$ is relevant for the mass generation of SM fermions while $\Phi_1$ is decoupled from the latter. The doublets can be written as,
\begin{equation}
\Phi _i = \begin{pmatrix} \phi ^+ _i \\ \left( v_i + \rho _i + i\eta _i \right)/ \sqrt{2}\end{pmatrix}
\end{equation}
with $v^2 = v_1^2 + v_2^2$ being the EW VEV.
The singlet scalar $\Phi_s= 1/\sqrt{2} \left( v_S + \rho _s + i \eta _s \right) $ is paramount to neutrino masses which are dynamically generated via the last two terms of Eq.\ (\ref{2hdm_tipoI_u1}) by the singlet VEV $v_S$. It leads to a mass matrix typical of the usual type I seesaw mechanism \cite{Mohapatra:1979ia,Mohapatra:1980yp,Schechter:1980gr},
\begin{equation}
\left(\nu \, N\right)
\left(\begin{array}{cc}
0 & m_D\\
m_D^T & M_R\\
\end{array}\right)\left(\begin{array}{c}
\nu \\
N \\
\end{array}\right),
\end{equation}
which results in  $m_\nu = -m_D^T/M_R m_D$ and $m_N = M_R$, as long as $M_R \gg m_D$, where $m_D= \frac{y^D v_2}{\sqrt{2}}$ and $M_R= \sqrt{2} y^M v_S$. In order to generate active neutrino masses at the sub-eV scale one can either set the right-handed neutrino masses at the  scale of a Grand Unified Theory (GUT) 
or else adopt suppressed Yukawa couplings \cite{BhupalDev:2012zg,Banerjee:2015gca}. The right-handed neutrinos, being charged under $U(1)_X$, may have a relevant impact on the collider bounds of the $Z^\prime$ because the latter can decay into them \cite{Emam:2007dy,Abdelalim:2014cxa,Accomando:2016sge,Accomando:2017qcs}. In this work we will conservatively assume all right-handed neutrinos to have the same mass of $100$~GeV. This assumption is conservative in the sense that, if the right-handed neutrinos were sufficiently heavy to prohibit the $Z^\prime$ boson to decaying into them, the Branching Ratio (BR) of the $Z^\prime$ into charged leptons or light quarks would be larger, thus strengthening our limits. \\

The scalar fields of the model are described by the following potential
\begin{equation}
\begin{split}
V &= m_{11} ^2 \Phi _1 ^\dagger \Phi _1 + m_{22} ^2 \Phi _2 ^\dagger \Phi _2 + \frac{\lambda _1}{2} \left( \Phi _1 ^\dagger \Phi _1 \right) ^2 + \frac{\lambda _2}{2} \left( \Phi _2 ^\dagger \Phi _2 \right) ^2  \\
&+ \lambda _3 \left( \Phi _1 ^\dagger \Phi _1 \right) \left( \Phi _2 ^\dagger \Phi _2 \right) + \lambda _4 \left( \Phi _1 ^\dagger \Phi _2 \right) \left( \Phi _2 ^\dagger \Phi _1 \right) \\
& + m_s ^2 \Phi _s ^\dagger \Phi _s + \frac{\lambda _s}{2} \left( \Phi _s ^\dagger \Phi _s \right) ^2 + \mu _1 \Phi _1 ^\dagger \Phi _1 \Phi _s ^\dagger \Phi _s \\
&+ \mu _2 \Phi _2 ^\dagger \Phi _2 \Phi _s ^\dagger \Phi _s + \left( \mu \Phi _1 ^\dagger \Phi _2 \Phi _s + h.c. \right).
\label{pot_2hdm_U1}
\end{split}
\end{equation}
Notice the absence in the potential, due to $U(1)_X$ invariance, of the $m_{12}^2 \Phi_1^\dagger \Phi_2$ quadratic term. In the standard $Z_2$ realization of the 2HDMs, one has to introduce an ad-hoc $m_{12}^2$ parameter that softly breaks the discrete symmetry. 
In these scenarios, instead, it is dynamically generated by the vev $v_S$ of the singlet scalar.
The presence of a scalar doublet charged under $U(1)_X$ (see table \ref{cargas_u1_2hdm_tipoI}) leads to $Z-Z^\prime$ mass mixing. This mass mixing is explicitly derived in {Appendix A}. Moreover, we also account for the presence of a kinetic mixing $\epsilon$ in the Lagrangian \cite{Babu:1997st,Langacker:2008yv,Gopalakrishna:2008dv,Bandyopadhyay:2018cwu,Abdullah:2018ykz}: 
\begin{equation}
\mathcal{L} _{\rm gauge} =  - \frac{1}{4} B _{\mu \nu} B^{\mu \nu} + \frac{\epsilon}{2\, \cos \theta_W} X _{\mu \nu} B^{\mu \nu} - \frac{1}{4} X _{\mu \nu} X ^{\mu \nu}. 
\label{Lgaugemix1}
\end{equation}
Here, $B$ and $X$ are the neutral vector bosons from the $U(1)_Y$ and $U(1)_X$ gauge groups, which, after EW Symmetry Breaking (EWSB) and together with the third component of the $SU(2)_L$ gauge bosons, give rise to the massless photon as well as massive $Z$ and $Z^\prime$.  \\

In summary, after taking into account the kinetic and mass mixings, we find the neutral current
\begin{widetext}
\begin{eqnarray}
\mathcal{L_{\rm NC}} \supset & - \left(\frac{g_Z}{2} J_{NC}^\mu \cos \xi \right) Z_\mu- \left(\frac{g_Z}{2} J_{NC}^\mu \sin \xi \right) Z^\prime_\mu \\
& + \frac{1}{4} g_X \sin \xi \left[ \left( Q_{Xf} ^R + Q_{Xf} ^L \right) \bar{\psi} _f \gamma ^\mu \psi _f + \left( Q_{Xf} ^R - Q_{Xf} ^L \right) \bar{\psi} _f \gamma ^\mu \gamma _5 \psi _f \right] Z_\mu \\
& - \frac{1}{4} g_X \cos \xi \left[ \left( Q_{Xf} ^R + Q_{Xf} ^L \right) \bar{\psi} _f \gamma ^\mu \psi _f - \left( Q_{Xf} ^L - Q_{Xf} ^R \right) \bar{\psi} _f \gamma ^\mu \gamma _5 \psi _f \right] Z' _\mu,
\label{LNC1}
\end{eqnarray}
\end{widetext}where $\xi$ is the mixing parameter explicitly given in  Eq.\ (\ref{eqsinxi}) while $Q^L_X$ ($Q^R_X$) are the left-handed (right-handed) fermion charges under $U(1)_X$ defined according to table~\ref{cargas_u1_2hdm_tipoI}. One can then easily obtain the $Z^\prime$ interactions with the SM fermions by substituting their charges for each of the models exhibited in table~\ref{cargas_u1_2hdm_tipoI}. This interaction Lagrangian represents the key information for the  collider phenomenology we are going to tackle, because it dictates  $Z^\prime$ production as well as its most prominent decays to be searched for. In principle, there are other interactions involving the $Z^\prime$ gauge boson besides the neutral current of Eq. \ref{LNC1}. For example, the mass and kinetic mixing will generate trilinear terms such as $Z^\prime h Z$, $Z^\prime h A$ (with $h$ being a CP-even Higgs boson) and $Z^\prime W^+W^-$, but these are suppressed by the small values of the $Z-Z^\prime$ mixing, thus not changing the overall $Z^\prime$ decay pattern significantly. In fact, we explicitly checked that these couplings change the overall bounds on the $Z^\prime$ mass up to $5\%$ at the most.\\

Having described the model and the relevant interactions, we can now present the collider bounds on the various $Z'$ realization we presented.

\begin{figure}[t!]
\includegraphics[width=\columnwidth]{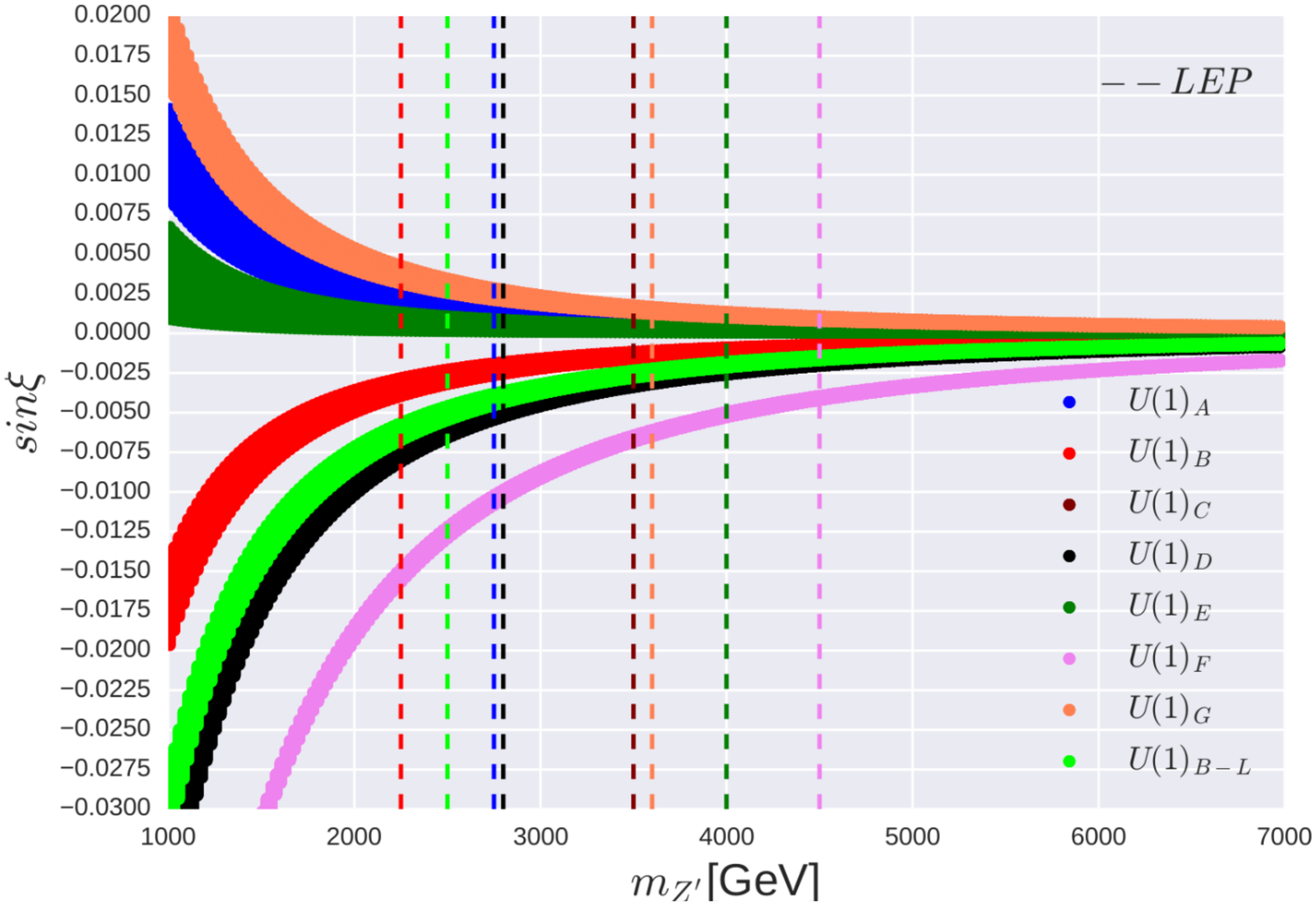}
\caption{LEP constraints (dashed vertical lines) in the plane $\sin \xi$ vs\ $m_{Z^\prime}$ for each of the $U(1)_X$ models considered in this work.}
\label{fig:LEP}
\end{figure}

\section{Collider Constraints}
\label{sec_collider}
The main goal of this section is to derive the LEP and LHC bounds on the $Z^\prime$ gauge bosons with charge assignments displayed in table \ref{cargas_u1_2hdm_tipoI}. We start by using LEP data \cite{LEP:2003aa} to constrain the kinetic and mass mixing terms which may significantly affect the $Z$ properties if the mixing angle ($\xi$) is sufficiently large. \\

Typically this mixing angle is assumed to be arbitrarily small in models where there is no tree-level kinetic and mass mixing 
\cite{Martinez:2014rea,Martinez:2014ova,Martinez:2015wrp,Okada:2016tci,Okada:2016gsh,Arcadi:2017kky,Arcadi:2018tly} 
but this assumption no longer applies to our models because we do include the kinetic mixing and the Higgs doublets are charged under the new gauge group (which implies the existence of mass mixing). In other words, once the gauge symmetry and the spontaneous EWSB mechanism are established, there is not much freedom left concerning the mass mixing, which is essentially set by the charges of the scalar fields under $U(1)_X$ and their VEVs. Moreover, the kinetic mixing, which can in principle be made ad hoc small, still arises at one-loop level since the SM fermions are charged under $U(1)_X$ \cite{Davoudiasl:2015hxa,Duerr:2016tmh,Aaij:2017rft}. 
After the analysis of the LEP constraints, we will then consider LHC bounds for generic Abelian extensions of the SM which, in the presence of sizeable kinetic mixing, have previously been obtained in \cite{Accomando:2016sge,Accomando:2016rpc,Accomando:2017qcs}. \\

\subsection{LEP Limits}

The LEP constraints on our models arise in the light of the excellent precision achieved by data collected and analyzed at such a machine. For example, by measuring processes such as $e^+e^- \rightarrow l^+ l-$, where $l=e,\mu$, LEP can restrictively probe beyond the SM scenarios that feature new particles coupling to charged leptons. In this connection, for our models, the presence of a massive $Z^\prime$ that mixes with the $Z$ leads to a deviation from the universal interactions of the latter with electrons and muons. This deviation can be parametrized as
\begin{eqnarray}
\delta\Gamma_{\mu e}=1-\frac{\Gamma(Z\rightarrow e^-e^+)}{\Gamma(Z\rightarrow \mu^-\mu^+)}. 
\label{eqGamma}
\end{eqnarray}
The parameter $\delta\Gamma_{\mu e}$ was measured at LEP \cite{LEP:2003aa} and can be presently used to place limits on the models studied in this work. Indeed the $Z-Z^\prime$ kinetic and mass mixing will make this quantity depart from unit by inducing new interactions between the $Z$ and SM fermions that are proportional to the mixing angle $\xi$, according to Eq. (8). Therefore, one can constrain such mixing using the LEP measurement of $\delta\Gamma_{\mu e}$ as shown in figure \ref{fig:LEP} by the dashed vertical lines. This idea was similarly exploited in the past to constrain different $Z^\prime$ models \cite{Leike:1992uf,Montagna:1995ty,Montagna:1996ie,Chiappetta:1996km,Richard:2003vc,Feldman:2007wj,Hayden:2013sra,Kim:2016bdu}.\\

We derived constraints on the mixing angle by plugging Eq. (8) into Eq. (\ref{eqGamma}) and then comparing with the aforementioned LEP measurement \cite{LEP:2003aa}. In order to better understand these bounds, we recall that the mixing angle can be well approximated in the limit of large $m_{Z^\prime}$ by
\begin{equation}
\sin \xi \sim (G_{X_1} v_1^2 + G_{X_2} v_2^2 )/m_{Z^\prime}^2,
\label{Eqsinxi}
\end{equation}where $G_{X_i}$ are the couplings defined in Eq.  (\ref{GXeq}) and encompass the gauge group dependence of the scalar doublets.
Hence, using  Eq. (\ref{Eqsinxi}), we can relate the mixing angle $\xi$ to the $Z^\prime$ mass for all $U(1)_X$ models and thus find the lower mass bounds indicated by the vertical dashed lines in figure \ref{fig:LEP}.
\\

Once a gauge charge assignment is picked, $G_{X_i}$ is determined up to $g_X$. Then, if we assume  $\epsilon=10^{-3}$ and $\tan \beta=10$ (see Appendix A),  in agreement with the latest experimental constraints on the 2HDM \cite{Bernon:2015wef}, the (sine of the)  mixing angle $\sin\xi$, $g_X$ and $m_{Z^\prime}$ are directly connected to one another leaving, in the end, only two free parameters. Their relation is model dependent and for this reason there is a curve for every model in figure \ref{fig:LEP}. (The curve for model $C$ is not visible in figure \ref{fig:LEP} because it is hidden between the curves for models $D$ and $B-L$.) \\

Notice that the vertical lines correspond to the lower mass bounds on the $Z^\prime$ mass obtained by enforcing that all $Z$ properties, crucially including its mass measurement $m_Z = 91.1876\pm 0.0021$, remain in agreement with LEP data. 
We list these limits in table \ref{tabLEP} as a function of the mixing angle $\xi$ and $Z^\prime$ mass.

\begin{table}
{\bf LEP}\\
\begin{tabular}{|c|c|c|}
\hline
Model & Mixing & Lower Mass Bound\\
\hline
$U(1)_A$ & $|\sin \xi| < 0.00188$ & $m_{Z^\prime} > 2.7$~TeV\\
$U(1)_B$& $|\sin \xi| < 0.00281$ & $m_{Z^\prime} > 2.25$~TeV\\
$U(1)_C$ & $|\sin \xi| < 0.00264$ & $m_{Z^\prime} > 3.5$~TeV\\
$U(1)_D$ & $|\sin \xi| < 0.00471$ & $m_{Z^\prime} > 2.8$~TeV\\
$U(1)_E$ & $|\sin \xi| < 0.00041$& $m_{Z^\prime} > 4.0$~TeV\\
$U(1)_F$ & $|\sin \xi| < 0.00049$ & $m_{Z^\prime} > 4.25$~TeV\\
$U(1)_G$ & $|\sin \xi| < 0.00166$ & $m_{Z^\prime} > 3.6$~TeV \\
$U(1)_{B-L}$& $|\sin \xi| < 0.00198 $ & $m_{Z^\prime} > 2.5$~TeV\\
\hline
\end{tabular}
\caption{Summary of the LEP bounds on the (sine of the) mixing angle $\xi$ {vs} $m_{Z^\prime}$  for all the $U(1)_X$ models in table \ref{cargas_u1_2hdm_tipoI}.}
\label{tabLEP}
\end{table}

It is now important to justify why our bounds do not explicitly depend on $g_X$. Indeed, in figure \ref{fig:LEP} our limits rely on the combination of $\sin\xi$ and $m_{Z^\prime}$ only. The dependence on $g_X$ enters in the $Z^\prime$ mass and in $\sin \xi$ via $G_{X_i}$, however, any change induced by $g_X$ can be parametrized as a shift on $v_S$ which is a free parameter and not an observable. In a nutshell, the $Z^\prime$ mass can be approximated for large values of $v_S$ as
\begin{equation}
m_{Z^\prime}\sim\frac{1}{2}q_X g_X v_S,
\label{Zpmass}
\end{equation} where $q_X$ is the $U(1)_X$ charge of the SM singlet scalar. Then any change on the LEP limits due to $g_X$ can absorbed back by a redefinition of $v_S$. \\
From table \ref{tabLEP} we conclude that $Z^\prime$ masses below $2$~TeV are excluded by LEP for all  models. In particular, for the $U(1)_F$ model LEP imposes $m_{Z^\prime} > 4.25$ TeV. We emphasize here that these limits  result from  $Z-Z^\prime$ mixing effects and not from $Z^\prime$ production at LEP. In {Appendix} A we also explicitly show how the $Z$ mass changes with the mixing angle and which scale of symmetry breaking, $v_S$, can be assumed in order to guarantee consistency with LEP data. Lower mass bounds of this nature are paramount for models that feature a large decay width lying outside the Narrow Width Approximation (NWA) which LHC limits are based on. We will come back to this point later on.\\

Anyhow, it is important to have these LEP bounds at hand since they already put strong limits on the mixing angles and on the $Z^\prime$ masses for each of the models discussed in this work. In the derivation of such bounds we made some assumptions on the kinetic mixing and $\tan\beta$. Nonetheless, the kinetic mixing is not much relevant because it only acts as a correction to the gauge coupling $g_X$, see Eq.\ref{GXeq}, which, as already explained above, can be absorbed by a rescaling of $v_S$. As for the value of $\tan \beta$ that enters in the $Z^\prime$ mass, its impact is not important either because $m_{Z^\prime}$ is mainly set by $v_S$ in the limit $v_S \gg v$.\\


Now that we have shown the LEP limits on our models we are ready to derive the LHC bounds.

\subsection{LHC Limits}
The best LHC bounds are obtained here by simulating at $13$~TeV CM energy the process
\begin{equation}
pp\rightarrow f\bar f +X \,,
\end{equation}where $f=q, l$, leading to dijet or dilepton signals, respectively. 
Since this channel is mediated by a heavy $Z^\prime$, alongside $\gamma$ and $Z$, a peak around the $Z^\prime$ mass would appear at large values of the invariant mass of the dijet or dilepton final state. Initially, we will describe this in our simulation by adopting the aforementioned  NWA, wherein the Breit-Wigner (BW) distribution capturing the propagation of the $Z^\prime$ is replaced by a Dirac $\delta$ distribution. Eventually, we will allow for finite width effects as well. In the Monte Carlo (MC) generation we also need to account for the possible presence of up to two extra jets from QCD radiation alongside $f\bar f$ production, which are represented by $X$ in the equation above. This obviously results into a better estimation of the production and decay cross section and its kinematics. What characterizes a dijet or dilepton signal in our study is of course the presence of an isolated jet or lepton pair with a large invariant mass. The numerical analysis is performed according to \cite{Aaboud:2017buh}. \\

Signals with a resonant peak at high dijet or dilepton masses are of course absent in the SM model thus making the observation of such events a smoking gun signature for new physics, especially for a $Z^\prime$ state. The main SM background stems from irreducible dijet and dilepton production via the $\gamma$ and $Z$ bosons, reducible $t\bar{t}$ production and decay as well as instrumental jet mis-reconstruction, but for invariant masses above $1$~TeV they total a few events only for, e.g., a Center-of-Mass (CM) energy of $8$~TeV and $\sim 20$ fb$^{-1}$ of integrated-luminosity \cite{Khachatryan:2014fba}.\\

In our work, firstly we aim at deriving LHC limits based on the dijet analysis performed by CMS with 13 TeV CM energy and $\mathcal{L}=12 $ fb$^{-1}$ \cite{Sirunyan:2016iap} as well as the dilepton study conducted by ATLAS with $\mathcal{L}=36 $ fb$^{-1}$ \cite{Aaboud:2017buh}, which are the corresponding most recent analyses for these datasets. To do so, we implemented all our $U(1)_X$ models in FeynRules \cite{Alloul:2013bka} and simulated the partonic events with MadGraph5 \cite{Alwall:2011uj}. We took into account hadronization and detector effects using Pythia8 \cite{Sjostrand:2007gs} and Delphes \cite{deFavereau:2013fsa}, respectively, with the so-called $k_T$-MLM jet matching scheme described in \cite{Mangano:2006rw}. \\

Since we are now discussing the on-shell production of a $Z^\prime$ gauge boson (i.e., in NWA), the key quantities are: (i) the $g_X$ coupling that enters both in the production cross section and $Z^\prime$ total width; the $Z^\prime$ BR into (ii) light quarks and (iii)  charged leptons; (iv) the $Z^\prime$ mass; (v) the  angle $\xi$ that controls the $Z-Z^\prime$ mixing. Notice that the latter is theoretically derived using Eq. (\ref{Eqsinxi}) once $g_X$ and the $Z^\prime$ mass are fixed. The fact that $\xi$ is directly fixed by the model parameters makes our study more predictive. While its value is taken compliant with LEP data in all cases, we note that the LHC offers orthogonal and independent constraints on the models. We need to assess, however, which LHC experiment provides the most stringent bounds. \\

\begin{figure*}[t!]
\includegraphics[width=\columnwidth]{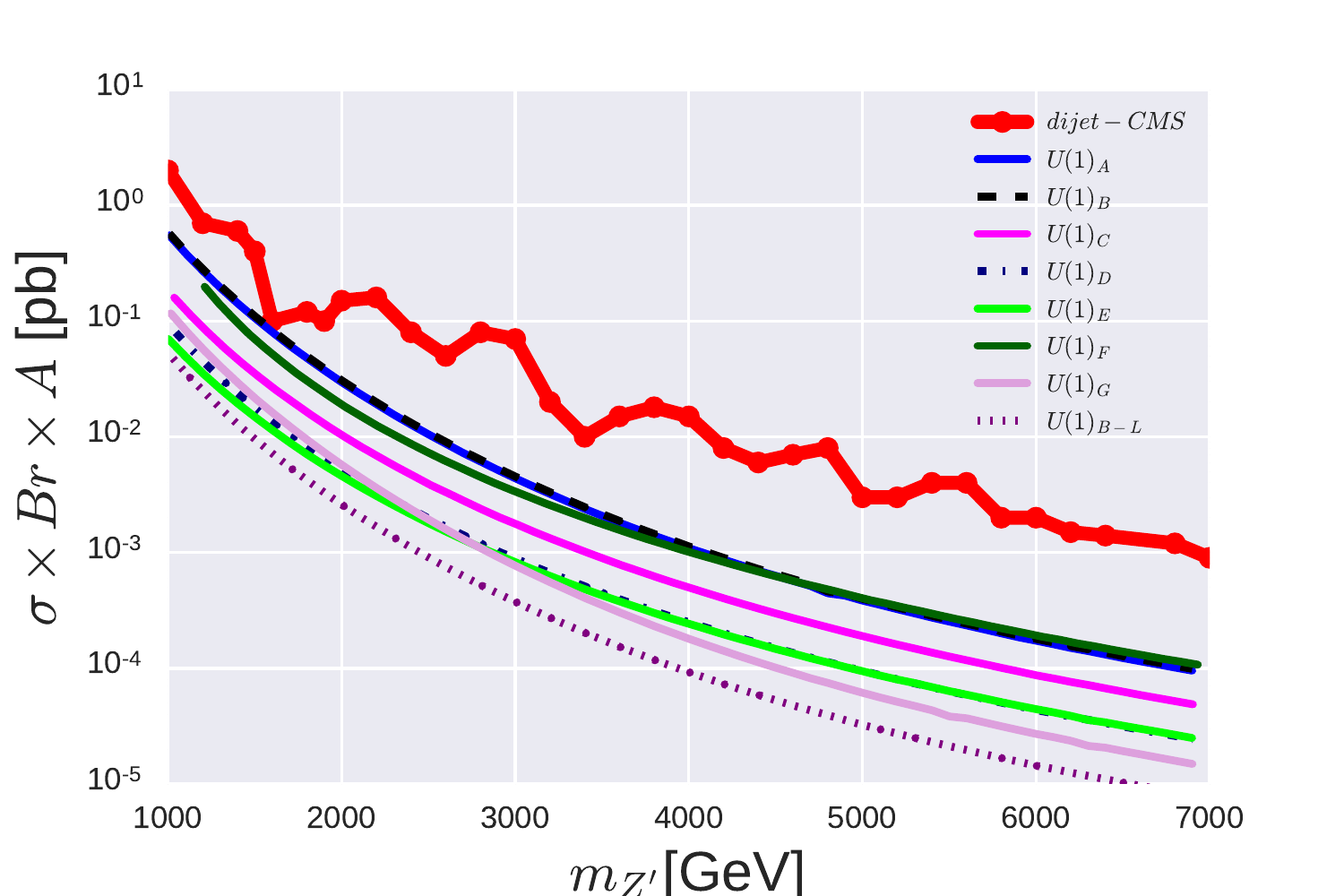}
\includegraphics[width=\columnwidth]{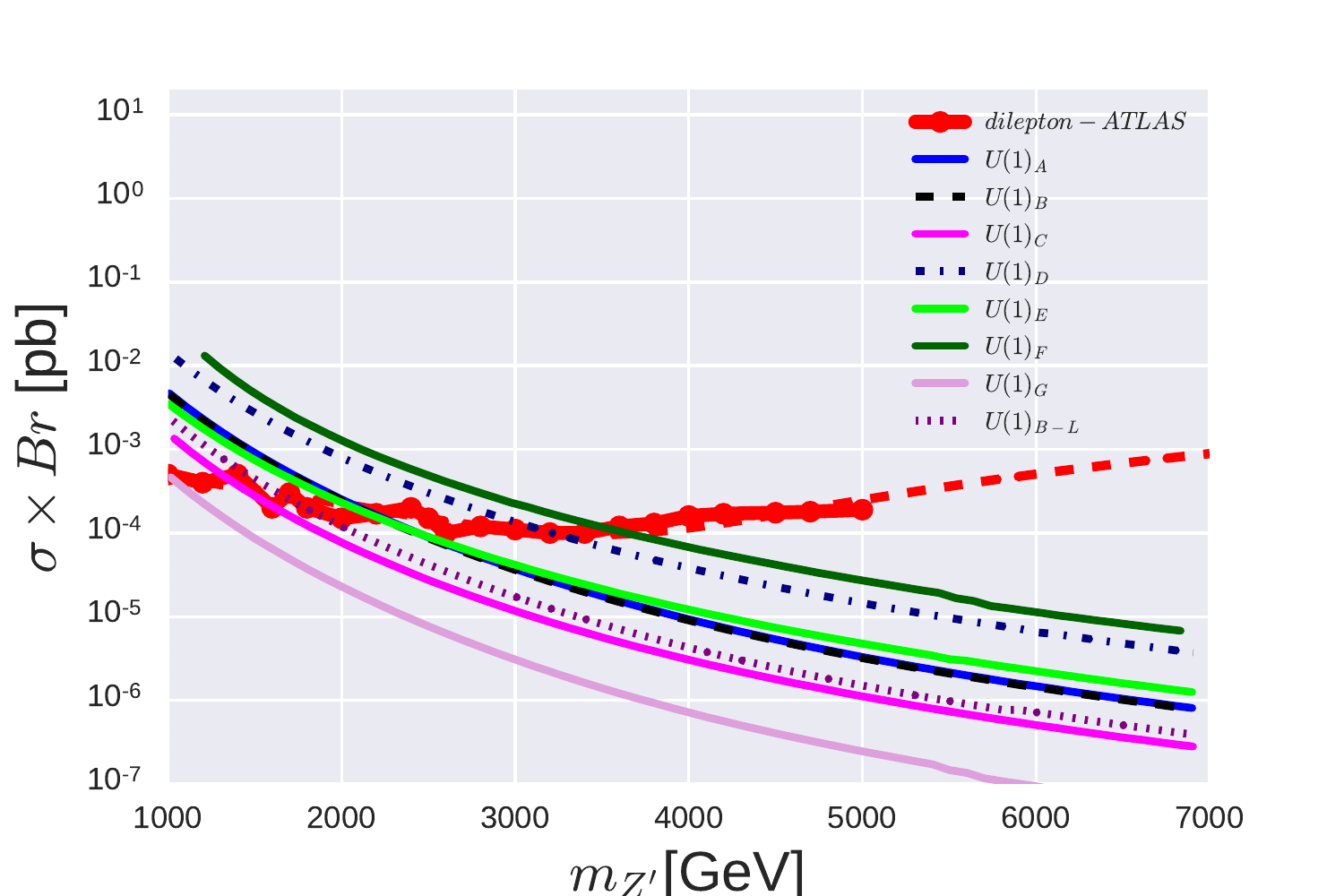}
\caption{ {Left panel:} The solid red curve represents the observed 95\% CL upper limit on the $Z^\prime $ production cross section times BR times acceptance $A$ for dijet events, where $A=0.6$ (data are taken from Ref. \cite{Sirunyan:2016iap}).  {Right panel}: The solid red curve represents the observed 95\% CL upper limit on the $Z^\prime $ production cross section times BR for dilepton events (data are taken from Ref. \cite{Aaboud:2017buh}). The other curves account  for the theoretical predictions as a function of $m_{Z^\prime}$ for all models discussed in this paper. The red dashed line in right panel represents the polynomial fit of the experimental data employed to extrapolate the ATLAS constraints up to 7 TeV. Here, $g_X=0.1$.}
\label{fig:dijet} 
\end{figure*}

Regarding the $Z^\prime$ BRs into light quarks and charged leptons, these change depending on the model under study because the fermions may have different quantum charges under $U(1)_X$. The $Z^\prime$ gauge boson features, in general, sizeable couplings to SM fermions making dijet and dilepton searches important environments to probe our models. Using the charge assignments defined above, we can therefore find lower bounds on the $Z^\prime$ mass by comparing our predictions with the experimental limits on the overall production and decay rates.

\begin{table}[!t]
\centering
LHC bounds for $g_X=0.1$\\
\begin{tabular}{|cc|}
\hline 
Models & Dilepton  \\ \hline 
$U(1)_A$ & $m_{Z^\prime}>2.2$~TeV \\
$U(1)_B$ & $m_{Z^\prime}>2.2$~TeV \\
$U(1)_C$ & $m_{Z^\prime} > 1.6$~TeV \\
$U(1)_D$ & $m_{Z^\prime}>3.5$~TeV  \\
$U(1)_E$ & $m_{Z^\prime}>2.3$~TeV \\
$U(1)_F$ & $m_{Z^\prime}>3.6$~TeV \\
$U(1)_G$ & $m_{Z^\prime} > 1.1$~TeV  \\
$U(1)_{B-L}$ & $m_{Z^\prime} > 2$~TeV  \\
\hline
\end{tabular}
\caption{Dilepton bounds on our 2HDMs with $U(1)_X$ gauge symmetries at 13 TeV CM energy using $\mathcal{L}=36 $ fb$^{-1}$ for $g_X=0.1$. The dijet limits for $g_X=0.1$ are weaker in comparison  thus are not displayed.}
\label{tabbounds1}
\end{table}

The LHC exclusion plots in NWA in the aforementioned two channels are presented in figure \ref{fig:dijet}. The CMS collaboration reported their limits in terms of the production cross section times BR into jets times acceptance (i.e., $\sigma\times {\rm BR} \times A$, where $A=0.6$ \cite{Sirunyan:2016iap}) whereas ATLAS provides the bounds on the production cross section times BR into charged leptons only (i.e., $\sigma\times BR)$ \cite{Aaboud:2017buh}. We have also extrapolated the ATLAS experimental bound from 5 to 7 TeV making a least-squares polynomial fit on the ATLAS data. Since at very large invariant masses errors are dominated by statistics, we can expect the experimental limit to indeed behave as reported on the right-hand side of figure \ref{fig:dijet}, but we emphasize here that this extrapolation is not robust despite it provides a reasonable estimate of the ATLAS bound for $Z^\prime$ masses above $5$~TeV. Notice that, by comparing the two panels in figure \ref{fig:dijet}, the dilepton limit is represented by a much smoother curve whereas the dijet one is rather bumpy due to the poorer reconstruction efficiency in the latter case. The first observation that we can easily draw by comparing the two plots in figure \ref{fig:dijet} is that dilepton bounds are more restrictive, as expected\footnote{Therefore, henceforth, we will no longer consider dijet data.}.

These limits are displayed in table \ref{tabbounds1} for all models considered. We highlight that all these bounds are derived with $g_X=0.1$, i.e, of EW strength.

As we pointed out above, each $Z'$ couples differently to the SM fermions making then the total width $\Gamma_{Z'}$ a model-dependent parameter.
As $Z'$ couplings to fermions grow (i.e., $g_X$ increases), $\Gamma_{Z'}$ also does, more so for large values of the $Z'$ mass, because of phase space effects. However, the relevant quantity playing a key role phenomenologically is the ratio between these two quantities, $\Gamma_{Z'}/m_{Z'}$.  
From now on, we will refer as NWA for the cases that accomplish $\Gamma_{Z'}/m_{Z'}\sim 1\%$ and as  Finite Width (FW) regime for the cases for which $\Gamma_{Z'}/m_{Z'}\sim 10\%$. It is important then to know when a particular model is in the NWA or FW setup. 

\begin{figure*}[t!]
\includegraphics[width=\columnwidth]{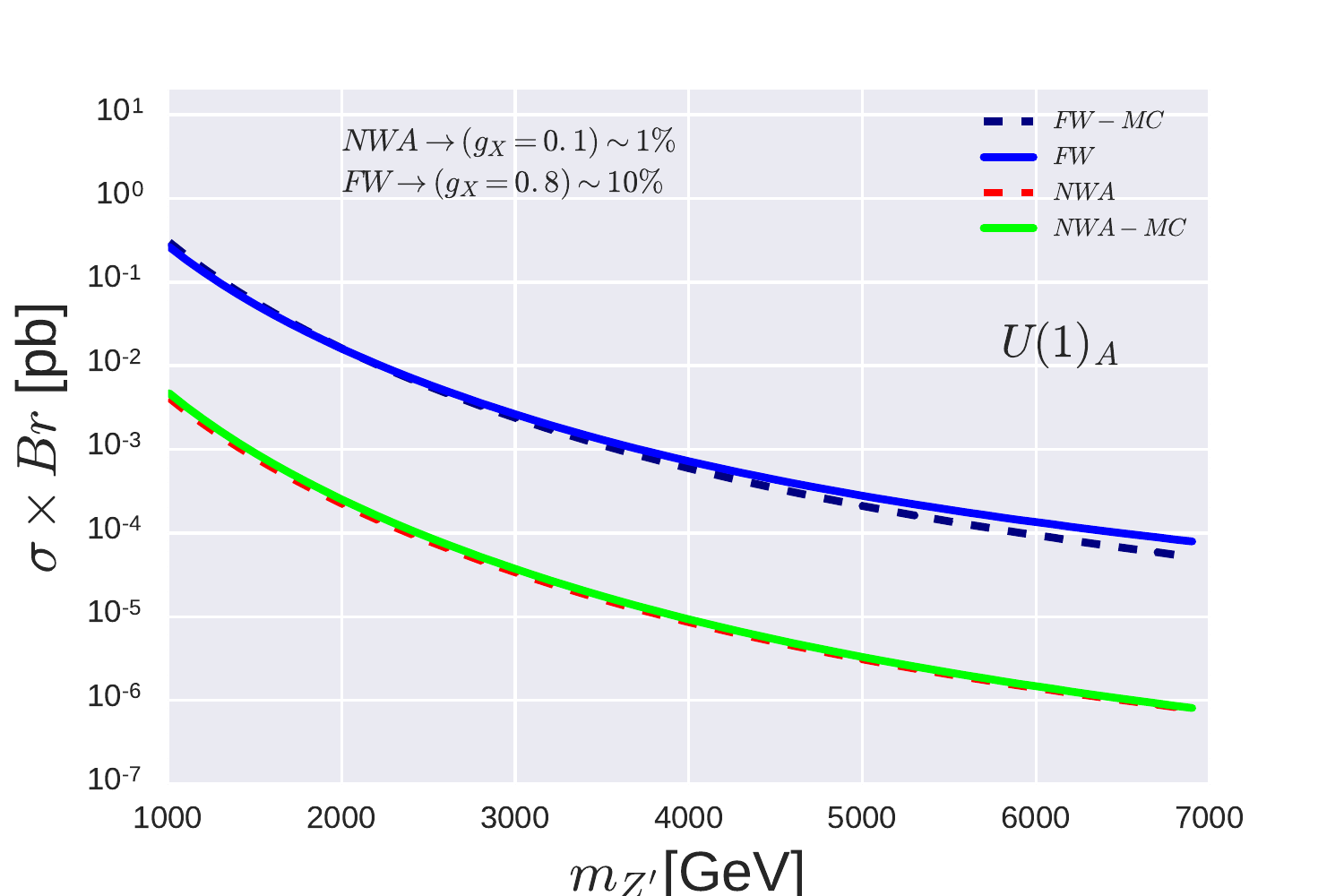}
\includegraphics[width=\columnwidth]{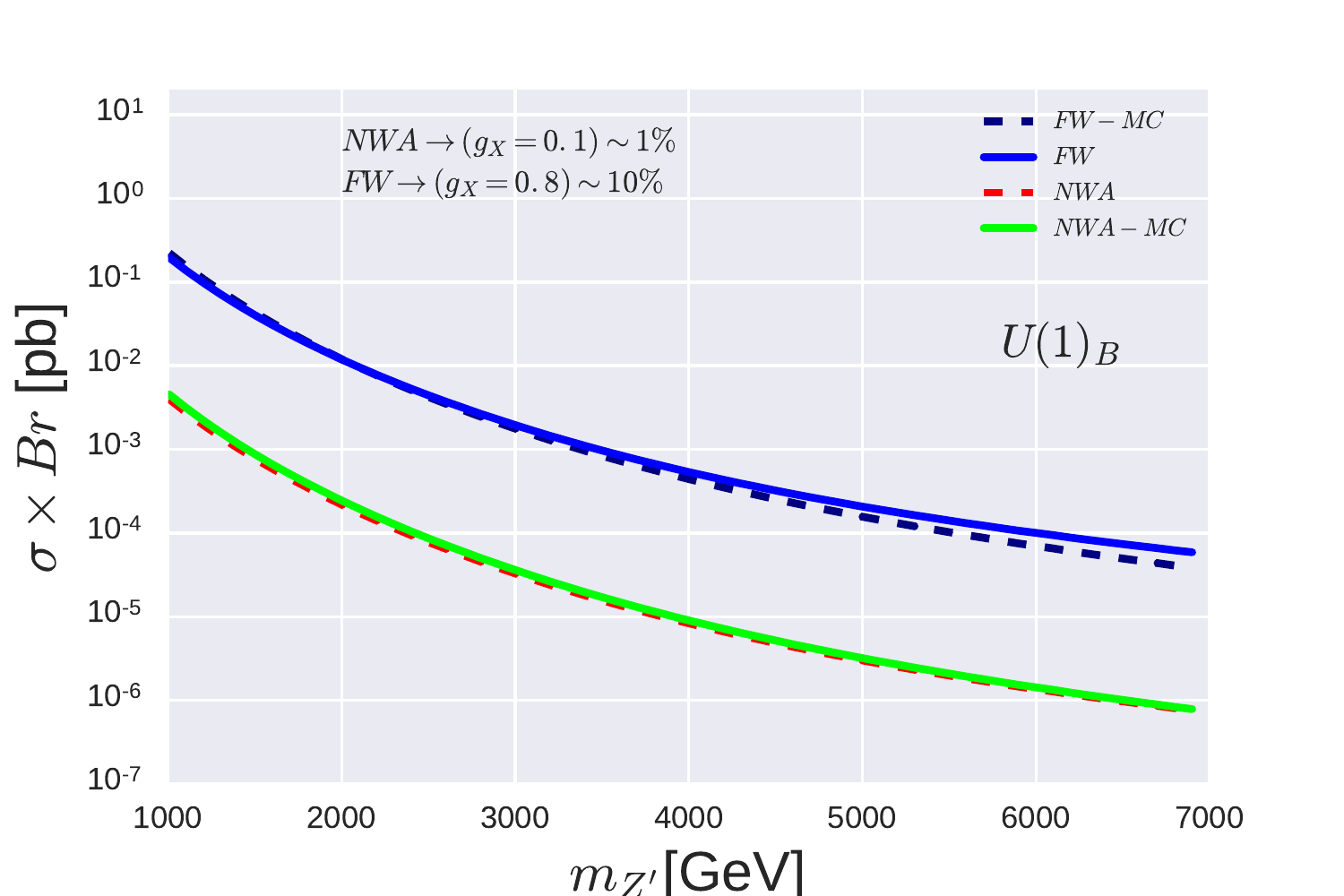}
\includegraphics[width=\columnwidth]{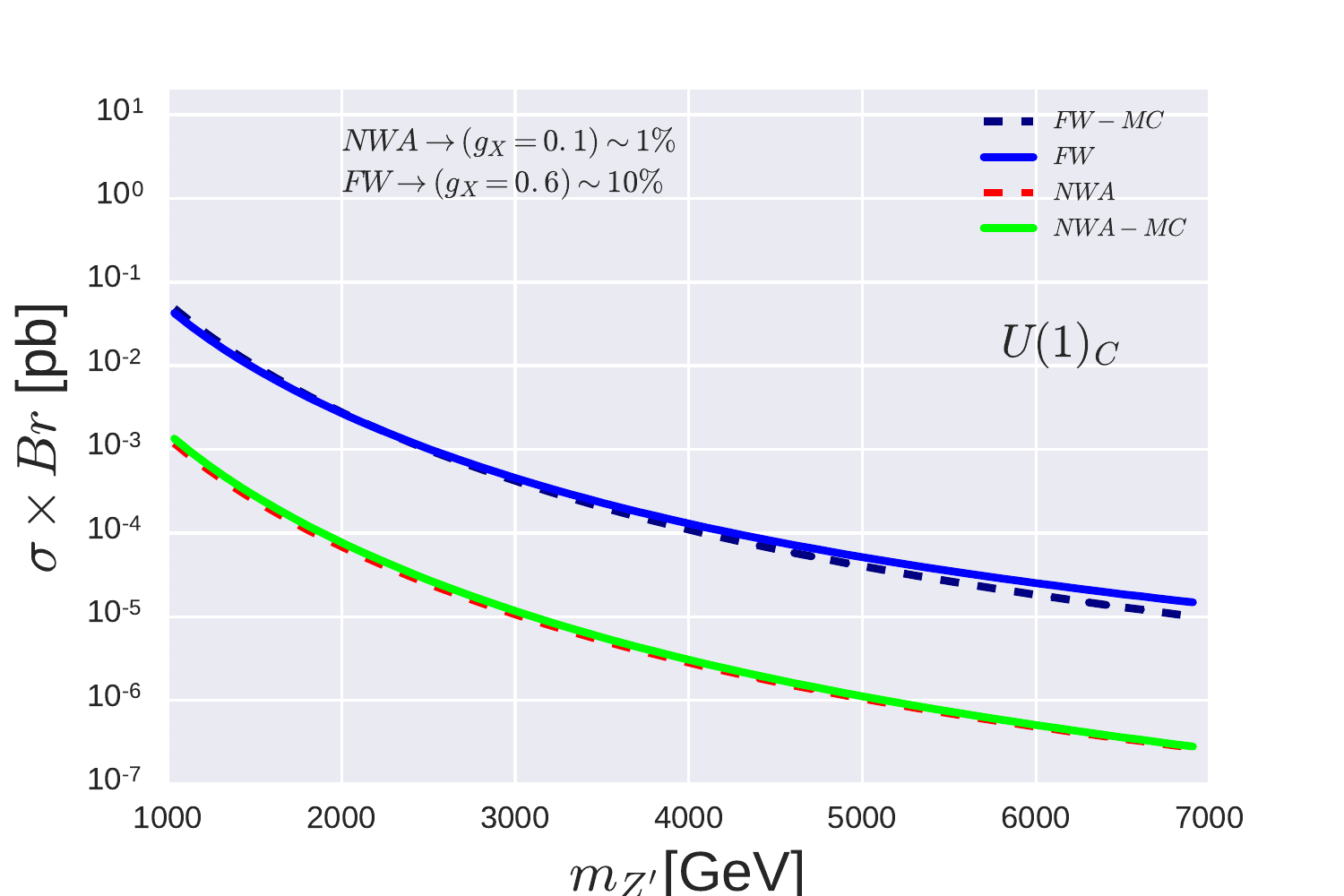}
\includegraphics[width=\columnwidth]{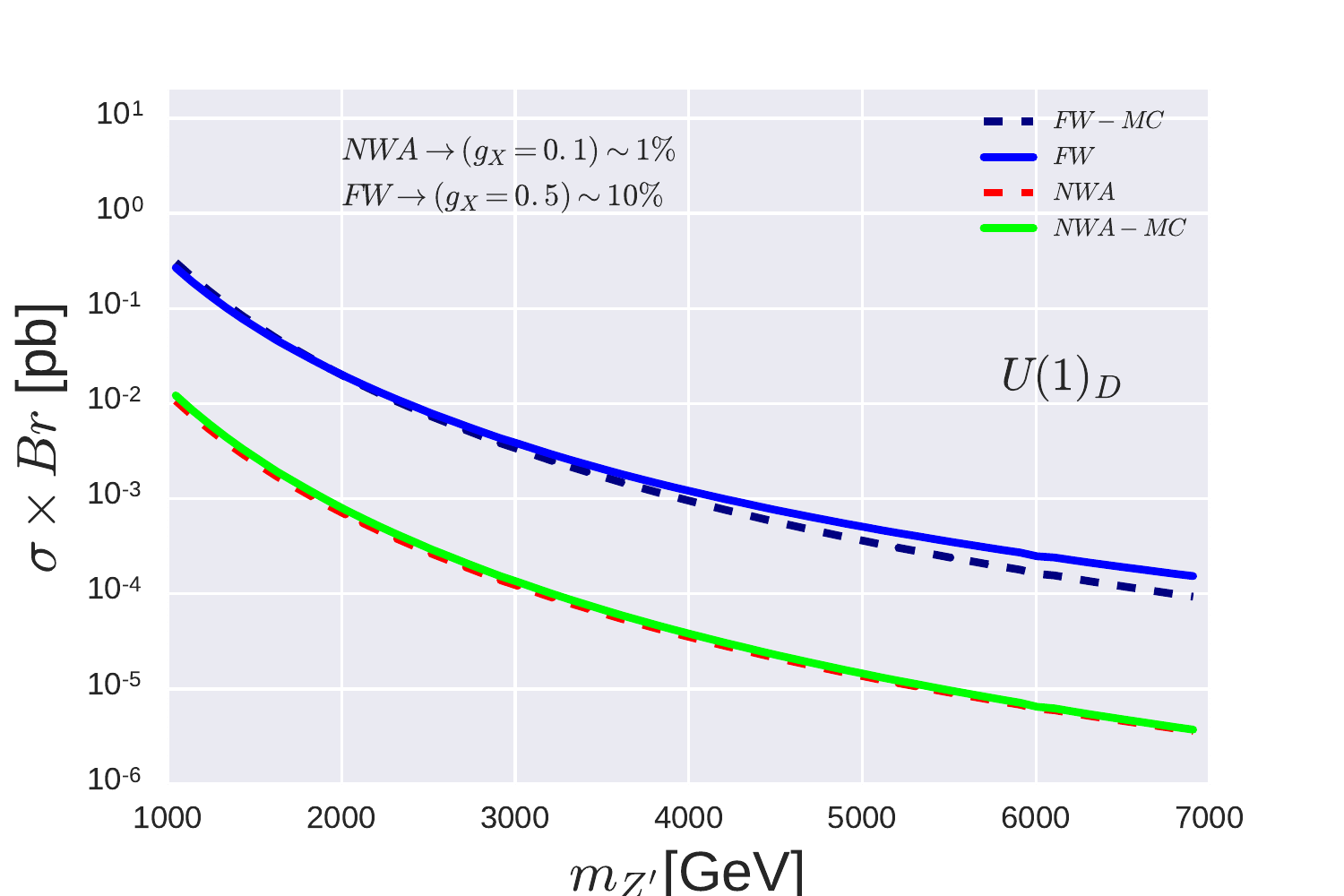}
\includegraphics[width=\columnwidth]{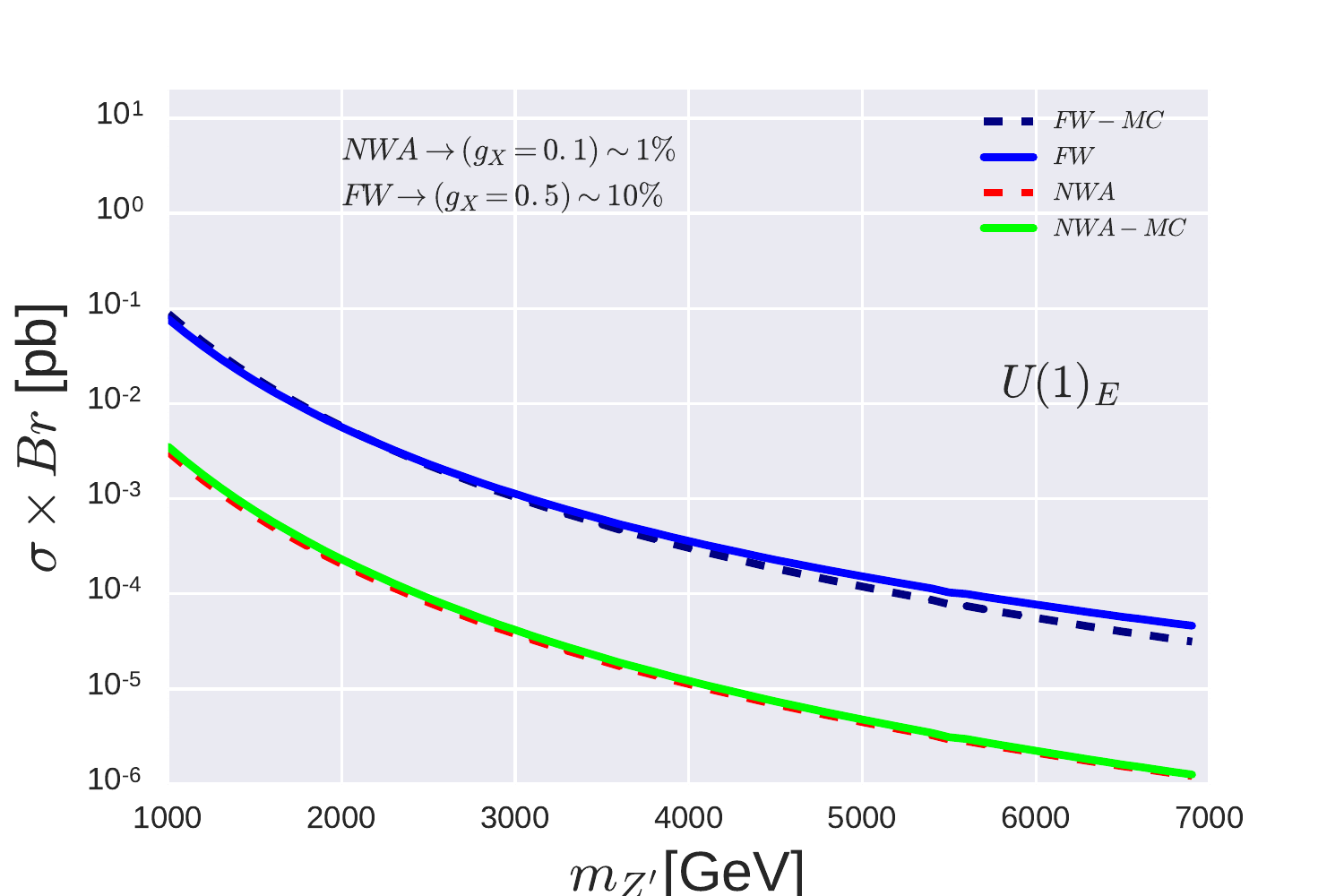}
\includegraphics[width=\columnwidth]{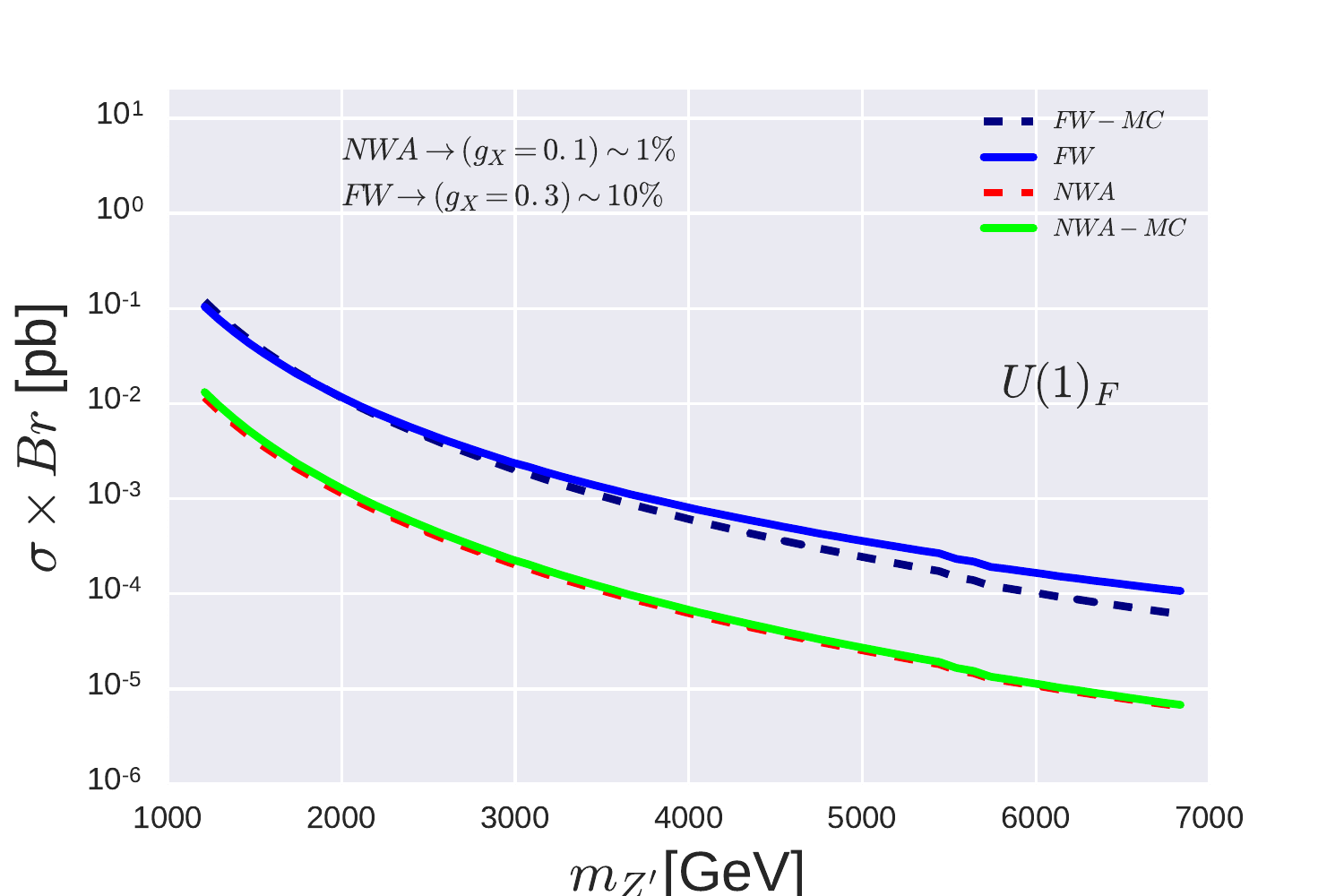}
\includegraphics[width=\columnwidth]{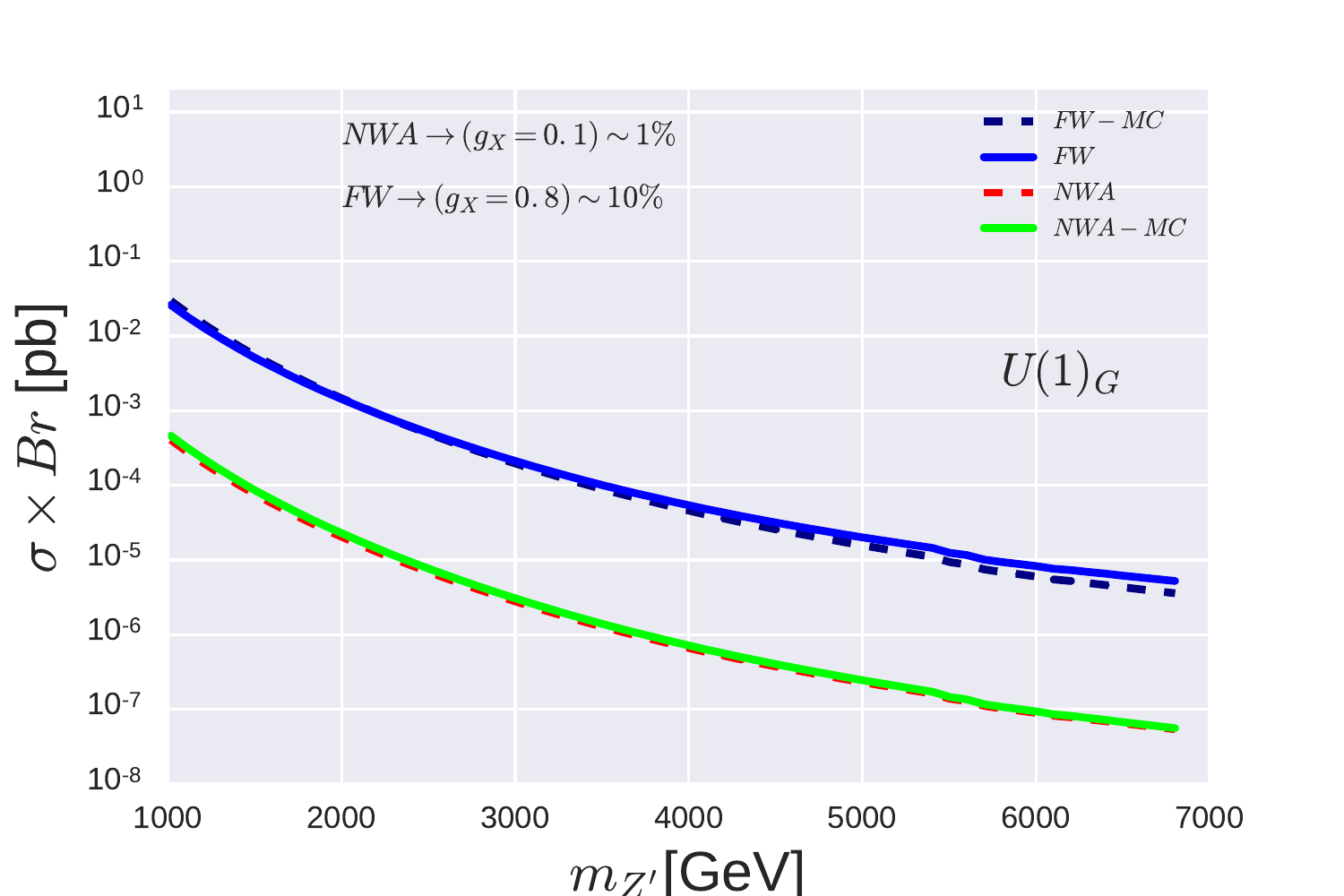}
\includegraphics[width=\columnwidth]{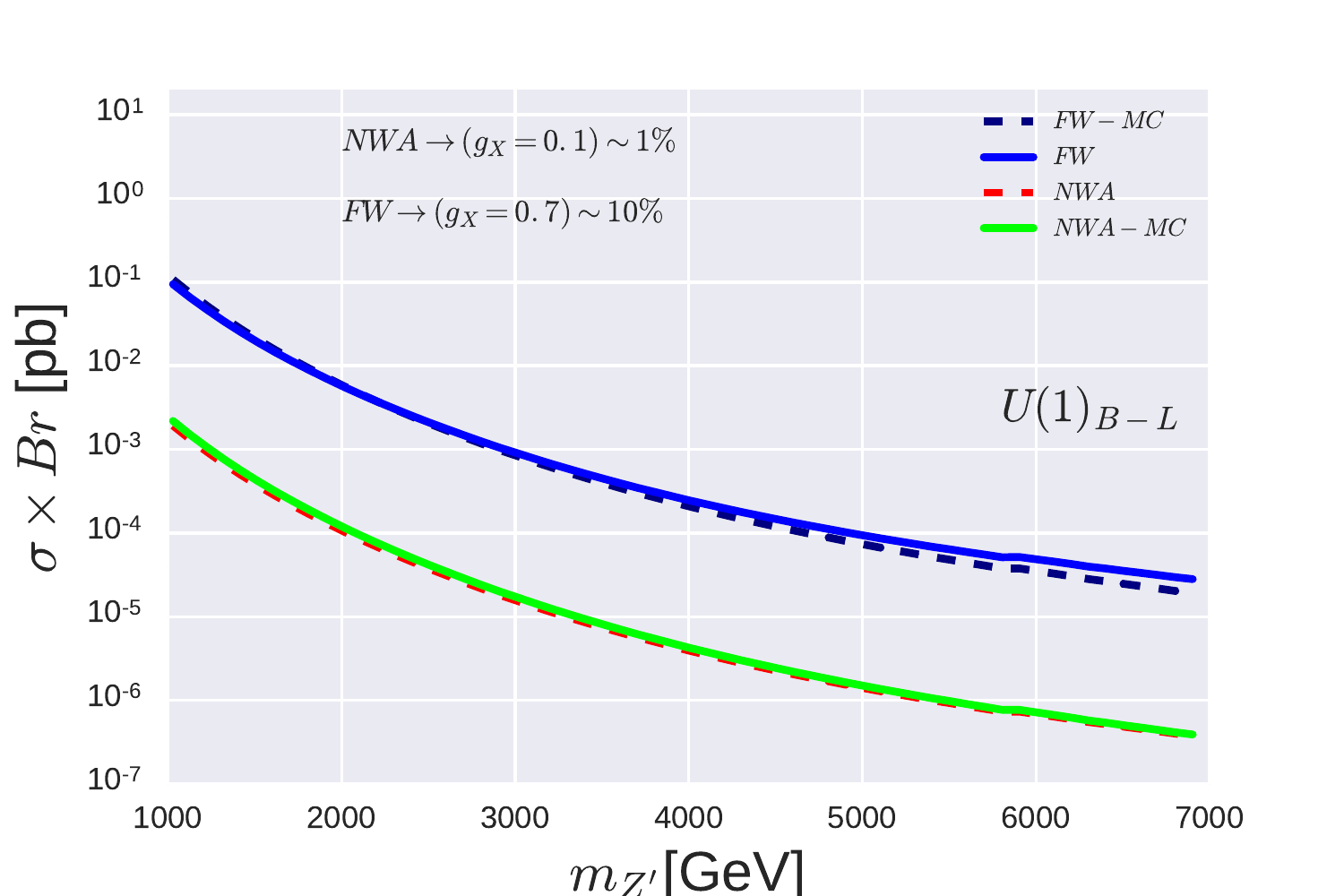}
\caption{Production cross section of a $Z^\prime$ times its dilepton BR  for each model considered assuming a NWA $(\Gamma_{Z'}/m_{Z'}\sim 1\%$) and FW regime ($\Gamma_{Z'}/m_{Z'}\sim \%10$). Alongside the rates without any cuts we also display those following the magic cut of 
\cite{Accomando:2013sfa}, i.e.,  
$|m_{ll}-m_{Z^\prime}|< 5\% \sqrt{s}$, which are denoted by the  NWA-MC and FW-MC labels.}
\label{fig:NWA_vs_FW}
\end{figure*}

In the Appendix we will show how $\Gamma_{Z'}$ changes for each model while $g_X$ varies from 0.1 to 1, indeed covering both cases, NWA and FW, for all  $U(1)_X$ models. Here, we have  calculated in figure \ref{fig:NWA_vs_FW}, for the BP with $\tan\beta =10$ and $\epsilon = 10^{-3}$, the single production cross section of $Z'$ times the dilepton BR for four cases of NWA and FW configurations, the ones without any cut on the dilepton invariant mass (NWA and FW) and the ones with the "magic cut" of 
\cite{Accomando:2013sfa}, i.e., 
$|m_{ll}-m_{Z^\prime}|< 5\% \sqrt{s}$ (NWA-MC and FW-MC), with $\sqrt{s}= 13$~TeV. Recall that Ref. \cite{Accomando:2013sfa} adopted such a constrain to capture interference effects between $Z'$ signals and corresponding SM irreducible backgrounds in such a way that these are minimized over the relevant kinematic range, thus enabling one 
to perform quasi-model-independent analyses, essentially preserving the NWA scheme most often used by the experimental collaborations for  $Z'$ boson searches.
Let us stress that each $U(1)_X$ model attains NWA or FW approximations for different values of $g_X$, for example, the $U(1)_G$ model achieve FW conditions for $g_X = 0.8$ while the $U(1)_F$ model does so for $g_X= 0.3$.  

\begin{figure*}[t!]
\includegraphics[width=\columnwidth]{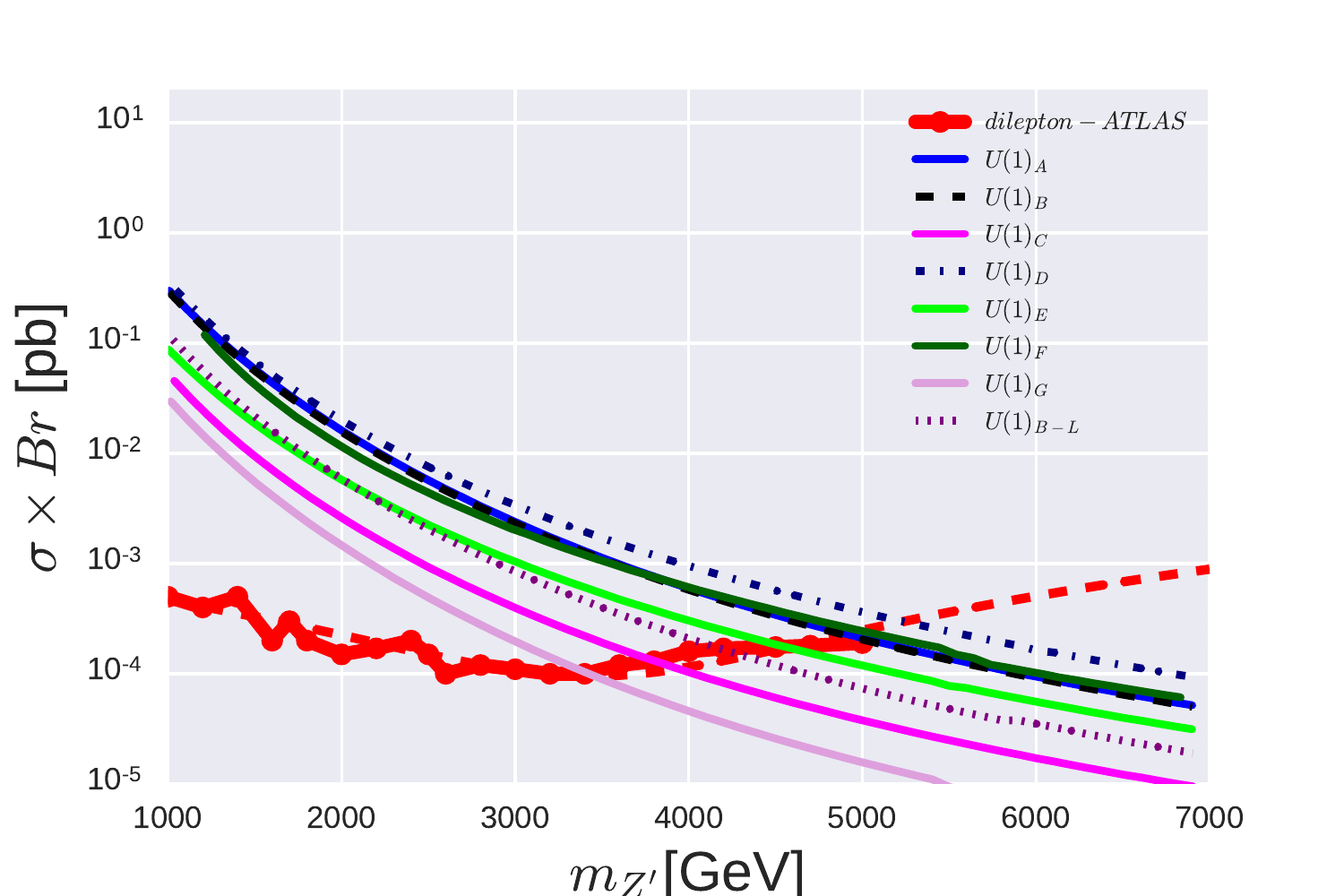}
\includegraphics[width=\columnwidth]{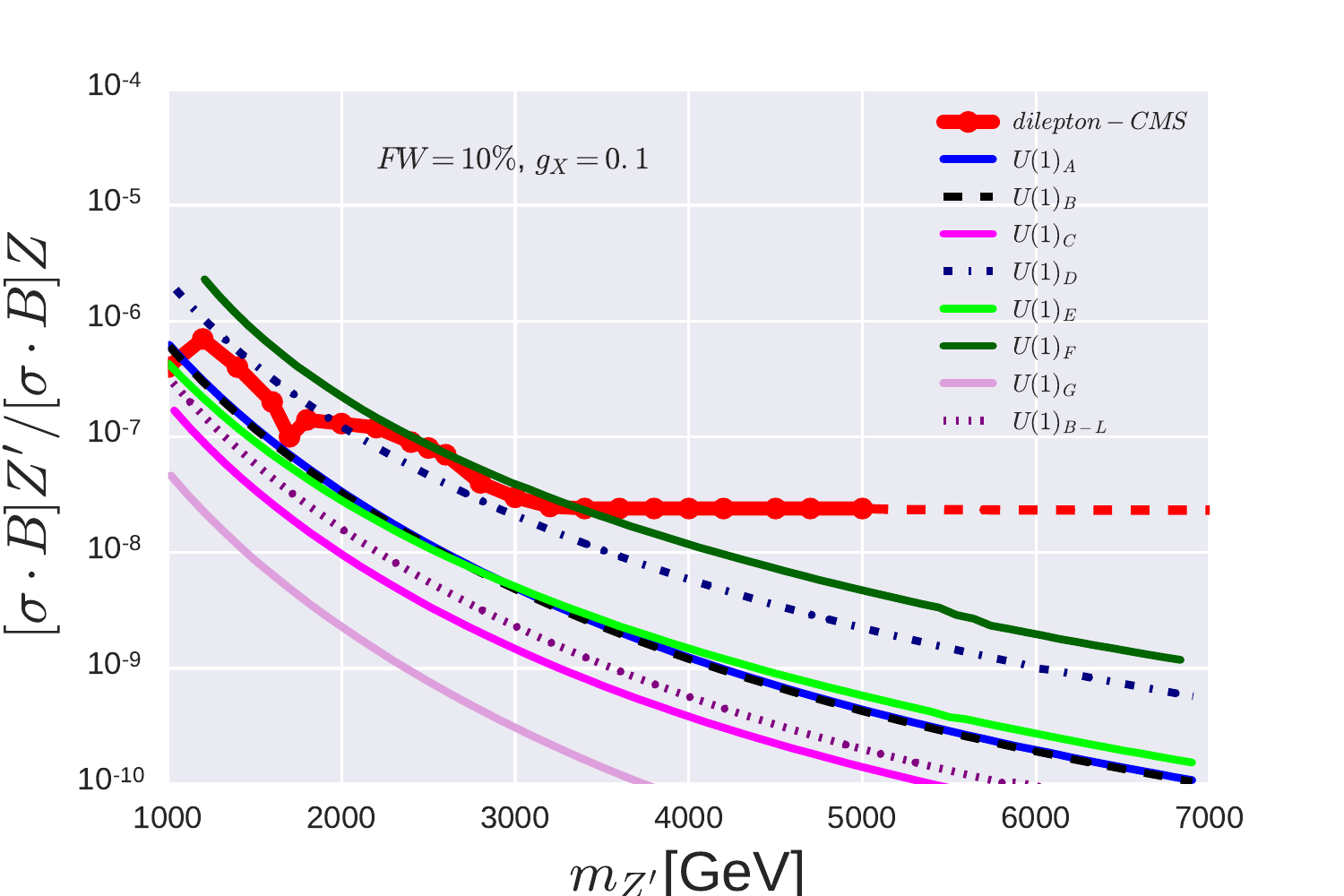}
\caption{{Left panel:} The solid red curve represents the observed 95\% CL upper limit on the $Z^\prime $ production cross section times BR for dilepton events (data are taken from Ref. \cite{Aaboud:2017buh}) while
the other curves are  for the theoretical predictions as a function of $m_{Z^\prime}$  taking different $g_X$ values yielding $\Gamma_{Z'}/m_{Z'}=10\%$  for all models considered. 
{Right panel}: The solid red curve represents the observed 95\% CL upper limit on the $Z^\prime $ production cross section times BR relative to the $Z$ production cross section times BR 
for dilepton events ({{data are taken from Ref. \cite{Sirunyan:2018exx}}}) while the other curves are for the theoretical predictions as a function of $m_{Z^\prime}$  taking  $g_X=0.1$ and setting by hand $\Gamma_{Z'}/m_{Z'}=10\%$ for all models considered. The red dashed line in both panels represents the polynomial fit of the experimental data employed to extrapolate the ATLAS and CMS constraints up to 7 TeV. Notice that the magic cut of 
\cite{Accomando:2013sfa}, i.e.,  $|m_{ll}-m_{Z^\prime}|< 5\% \sqrt{s}$ has been applied here on both CMS data and MC predictions (right panel only).}
\label{fig:dijet3} 
\end{figure*}

The first observation that we can draw from figure \ref{fig:NWA_vs_FW} is that the magic cut has no impact at all on the cases in NWA, as expected, and a rather moderate impact in the FW regime and only for large masses, where interference effects are more substantial, the effects being essentially the same for all models considered. In this respect, it is worth noting that, since the FW regime is achieved here by increasing the coupling $g_X$, this implies not only a larger width but also a much larger cross section. In fact, in this case, at large dilepton invariant mass, the resonant $Z'$ contribution is dominant over the interference between the $Z'$ contribution and the SM one due to $\gamma$ and $Z$ exchange, as generally $g_X> e$ and $e/\sin\theta_W$ plus the $Z'$ is resonant while $\gamma$ and $Z$ are highly off-shell. In these conditions, the NWA result and the one in FW are very similar, as already shown in Ref. \cite{Accomando:2010fz}. We can then use  
the ATLAS data of Ref. \cite{Aaboud:2017buh} which assume a narrow $Z'$, even when $\Gamma_{Z'}/m_{Z'}=10\%$, which is here obtained by suitably adjusting $g_X$ (differently) for all models considered in order to extract limits. We do so in the left frame of figure \ref{fig:dijet3}. We observe that in this FW case the excluded $m_{Z'}$ range is comparable to the one of the previous case in NWA, as expected.  

The magic cut becomes important when we obtain the prescribed $\Gamma_{Z'}/m_{Z'}=10\%$ ratio by enforcing the latter by hand while adopting $g_X=0.1$ throughout our models. This is well justified. Recall in fact that $U(1)_X$ models like those considered here can be remnants of Grand Unification Theories (GUTs), wherein there can be large particle spectra (of both fermions and bosons) into which such heavy $Z'$s can decay to, so as to justify a 10\% width-to-mass ratio. Besides, in experimental searches, mass and width are treated as independent parameters. In this case, interference effects are substantial so that the constraint advocated by Ref. \cite{Accomando:2013sfa} is mandatory. We thus need to use experimental data selected using the magic cut, which are those of Ref. \cite{Sirunyan:2018exx}. In this analysis, see figure 4 therein, CMS plots the upper limits at 95\% CL on the product of production cross section and BR
 for a $Z'$ with finite width, including the case of 10\% of the resonance mass, relative to the product of production cross section and BR for the $Z$ boson. Hence, we present the right frame of figure \ref{fig:dijet3}. We observe here that the discussed interference effect is substantial, as a much more restricted $m_{Z'}$ range can be probed with respect to the cases when this is negligible, signalling that the correction is predominantly negative.

We can now compare our LHC bounds with those from LEP and notice that they are rather complementary to each other. For instance, LHC requires $m_{Z^\prime} > 3.6$~TeV for $g_X=0.1$ for the $U(1)_F$ model even when away from the NWA, see {Appendix}, though this is subject to systematic errors (of model-dependent FW and interference effects that ought to be quantified for each energy and luminosity values). LEP instead sets a lower mass limit of $m_{Z^\prime} > 2.5$~TeV constituting a weaker but more robust bound on the $Z^\prime$ mass on this case. This clearly demonstrates that LEP data is still a powerful tool to constrain new physics models, especially  those that feature a large width-to-mass ratio. Such scenarios are commonly used in dark matter model building endeavors \cite{Arcadi:2017hfi}, for example. For model $U(1)_D$, in contrast, the $Z^\prime$ does have a narrow width for $g_X=0.1$, with LHC(LEP) excluding $Z^\prime$ masses below  $3.5$~TeV($2.8$~TeV). In this case, the simpler NWA analysis described above is applicable and the LHC searches provide not only the strongest bound on the $Z^\prime$ mass but also a very solid one. 

In summary, one needs to truly consider both LEP and LHC data to extract reliable bounds on the $Z^\prime$ mass and couplings of $U(1)_X$ models, bearing in mind that these limits are complementary and orthogonal to each other.

\section{HL-LHC and HE-LHC Sensitivity}
\label{sec_HL-HE-LHC}

Bearing in mind that the models studied in this work predict a large $Z^\prime$ width, relative to its mass, for $g_X \sim 1$ and that this may imply a reassessment of the validity of the magic cut in presence of much larger luminosities and/or energies of the LHC, including that of systematic effects from both the theoretical and experimental side,
we will derive the projected sensitivity for the HL-LHC and HE-LHC configurations only for $g_X=0.1$.

The HL-LHC setup is characterized by $\mathcal{L}=300$ and $3000$ fb$^{-1}$ while the HE-LHC represents the LHC upgrade phase with CM energy of $27$~TeV. To find their physics sensitivity we will adopt the strategy described in \cite{Papucci:2014rja}. In short, what the code does is to solve an equation for $M_{\rm new}$ (i.e., the new limit on $m_{Z'}$), knowing the current bound $M$, as follows:
\begin{equation}
\frac{ N_{\rm signal\, events} (M_{\rm new}^2, E_{\rm new},\mathcal{L}_{\rm new})}
     { N_{\rm signal\, events} (M^2, 13\,{\rm TeV},36 {\rm fb}^{-1})}=1,
\end{equation}
with obvious meaning of the subscripts.

The results from this iteration are summarized in table \ref{LHCforcast}. The lower mass bounds found presently compared to the expected ones at the HL-LHC and/or HE-LHC clearly show how important is any LHC upgrade to test new physics models including a $Z'$. For some models such as, e.g., the $U(1)_A$, the HE-LHC will potentially probe $Z^\prime$ masses up to $7$~TeV, i.e., three times higher than with current data. In summary, the HL-LHC and (especially) the HE-LHC will represent discovery machines being able to probe gauge boson masses with unprecedented sensitivity up to the $O(10~{\rm TeV})$ domain. \\

\begin{table*}[!t]
\centering
\begin{tabular}{ccccccccc}
\hline 
Model & $13$~TeV, 36 fb$^{-1}$ & $13$~TeV, 300 fb$^{-1}$ &  $13$~TeV, 3000 fb$^{-1}$ & $27$ TeV, 300 fb$^{-1}$  & $27$~TeV, 3000 fb$^{-1}$\\ \hline 
$U(1)_A$ & $2.2$~TeV & $3.07$~TeV & $4.09$~TeV & $5.02$~TeV & $7.03$~TeV \\
$U(1)_B$ & $2.2$~TeV & $3.07$~TeV & $4.09$~TeV & $5.02$~TeV & $7.03$~TeV \\
$U(1)_C$ & $1.6$~TeV & $2.37$~TeV & $3.34$~TeV & $3.73$~TeV & $5.54$~TeV \\
$U(1)_D$ & $3.5$~TeV & $4.45$~TeV & $5.46$~TeV & $7.76$~TeV & $9.89$~TeV \\
$U(1)_E$ & $2.3$~TeV & $3.18$~TeV & $4.21$~TeV & $5.24$~TeV & $7.27$~TeV \\
$U(1)_F$ & $3.6$~TeV & $4.55$~TeV & $5.56$~TeV & $7.97$~TeV & $10.09$~TeV \\
$U(1)_G$ & $1.1$~TeV & $1.73$~TeV & $2.60$~TeV & $2.62$~TeV & $4.16$~TeV \\
$U(1)_{B-L}$ & $2$~TeV & $2.84$~TeV & $3.85$~TeV & $4.60$~TeV  & $6.55$~TeV \\
\hline
\end{tabular}
\caption{HL-LHC and HE-LHC projected sensitivities for all $U(1)_X$ models studied in this work using dilepton data at $13$~TeV and $27$~TeV CM energy and for $\mathcal{L}=36$, 300 and 3000 fb$^{-1}$.}
\label{LHCforcast}
\end{table*}

\section{Conclusion}
\label{sec_conclusion}
2HDMs offer an interesting framework for Higgs boson phenomenology above and beyond what offered by the SM but, at the same time, also face the presence of dangerous FCNCs which are severely constrained by data. An ad-hoc $Z_2$ symmetry is usually imposed to prevent one of the two scalar doublets from generating fermions masses and thus freeing 2HDMs from these  constraints. In this work, we derived collider bounds on 2HDMs that instead address the flavor problem via gauge symmetries and naturally accommodate  neutrino masses via a type-I seesaw mechanism. These gauge symmetries give rise to $Z^\prime$ gauge bosons that feature different coupling structures to leptons and quarks. We thus exploited the complementarity between dijet and dilepton data from the LHC to find lower mass bounds on the corresponding $Z^\prime$ gauge bosons, by investigating  scenarios where  these are applicable in both NWA and FW regime, finally contrasting the ensuing limits extracted from LEP data which are sensitive, on the other hand, to the effects of the kinetic and mass $Z-Z^\prime$ mixings. 
Lastly, we have presented the sensitivities of the HL-LHC and HE-LHC to such new physics, showing that they are capable of probing $Z^\prime$ masses up to a factor three higher than presently.

. \acknowledgments
The authors thank Werner Rodejohann and Pyung-won Ko for discussions and comments. DC and FSQ acknowledge financial support from MEC and UFRN. FSQ also acknowledges the ICTP-SAIFR FAPESP grant 2016/01343-7 for additional financial support. SM is supported in part by the NExT Institute and acknowledges partial  financial support from the STFC Consolidated Grant ST/L000296/1
and the H2020-MSCA-RISE-2014 grant no. 645722 (NonMinimalHiggs).

\section*{Appendix A}\label{appendixA}
We collate here some information aiding the understanding of the main part of the paper, by dividing it into sections relating to key computational aspects.

\section*{Gauge Boson Couplings}
After rotating to a basis in which the gauge bosons have canonical kinetic terms, the covariant derivative in terms of small $\epsilon$ reads 
\begin{equation}
\label{der_cov_u1_diag}
\small
D_\mu = \partial _\mu + ig T^a W_\mu ^a + ig ' \frac{Q_Y}{2} B _{\mu} + \frac{i}{2} \left( g ' \frac{\epsilon Q_{Y}}{\cos \theta _W} + g_X Q_X \right) X_\mu ,
\end{equation}
or, explicitly,
\begin{equation}
\small
 D_\mu = \scalemath{0.8}{\partial _\mu + \frac{i}{2} \begin{pmatrix} g W_\mu ^3 + g ' Q_{Y} B_\mu + G_X X_\mu & g \sqrt{2} W_\mu ^+ \\ g \sqrt{2} W_\mu ^- & - g W_\mu ^3 + g ' Q_{Y} B_\mu + G_X X_\mu \end{pmatrix}} ,
\end{equation}
where we defined for simplicity 
\begin{equation}
\label{GXeq}
G_{Xi} = \dfrac{g ' \epsilon Q_{Y_i}}{\cos \theta _W} + g_X Q_{X_i}  , 
\end{equation}with $Q_{Y_i}$ being the hypercharge of the scalar doublet, which in the 2HDM is taken equal to $+1$ for both scalar doublets, and $Q_{X_i}$ is the charge of the scalar doublet $i$ under $U(1)_X$.\\

Then the part of the Lagrangian responsible for the gauge boson masses becomes 
\begin{equation}
\begin{split}
\mathcal{L} _{\text{mass}} = & \scalemath{0.8}{\left( D_\mu \Phi _1 \right) ^\dagger \left( D^\mu \Phi _1 \right) + \left( D_\mu \Phi _2 \right) ^\dagger \left( D^\mu \Phi _2 \right) + \left( D_\mu \Phi _S \right) ^\dagger \left( D^\mu \Phi _S \right)}\\
 = & \scalemath{0.8}{\frac{1}{4} g^2 v ^2 W_\mu ^- W ^{+ \mu} + \frac{1}{8} g_Z ^2 v ^2 Z_\mu ^0 Z^{0 \mu} - \frac{1}{4} g_Z \left( G_{X1} v_1 ^2 + G_{X2} v_2 ^2 \right) Z_\mu ^0 X ^\mu}  \\
&+ \frac{1}{8} \left( v_1 ^2 G_{X1} ^2 + v_2 ^2 G_{X2} ^2 + v_S ^2 g_X ^2 q_X ^2 \right) X_\mu X ^\mu,
\end{split}
\label{mixinggaugebosons}
\end{equation}
where $v ^2 = v_1 ^2 + v_2 ^2$. Eq. \eqref{mixinggaugebosons} can then be written as
\begin{equation}
\begin{split}
\mathcal{L} _{\rm mass} &= \scalemath{0.8}{m_W ^2 W_\mu ^- W ^{+ \mu} + \frac{1}{2} m_{Z^0} ^2 Z_\mu ^0 Z^{0 \mu} - \Delta ^2 Z_\mu ^0 X ^\mu + \frac{1}{2} m_X ^2 X_\mu X ^\mu},
\end{split}
\end{equation}
with
\begin{equation}
m_W ^2 = \frac{1}{4} g^2 v ^2, \qquad m_{Z} ^2 = \frac{1}{4} g_Z ^2 v ^2,
\label{Eq:MZ0}
\end{equation}
\begin{equation}
\Delta ^2 = \frac{1}{4} g_Z \left( G_{X1} v_1 ^2 + G_{X2} v_2 ^2 \right),
\label{Eq:Delta2}
\end{equation}
and
\begin{equation}
m_X ^2 = \frac{1}{4} \left( v_1 ^2 G_{X1} ^2 + v_2 ^2 G_{X2} ^2 + v_S ^2 g_X ^2 q_X ^2 \right) .
\label{Eq:MX}
\end{equation}
Summarizing, after the symmetry breaking, one can realize that there is a remaining mixing between $Z^0 _\mu$ and $X_\mu$ that can be expressed through the symmetric matrix
\begin{equation}
m_{Z^0X} ^2 = \frac{1}{2} \begin{pmatrix} m_{Z^0} ^2 & - \Delta ^2 \\ \cdot & m_X ^2 \end{pmatrix}, 
\end{equation}
or, explicitly,
\begin{equation}
m_{Z^0X} ^2 = 
\frac{1}{8} \begin{pmatrix} g_Z ^2 v ^2 & - g_Z \left( G_{X1} v_1 ^2 + G_{X2} v_2 ^2 \right) \\ 0  & v_1 ^2 G_{X1} ^2 + v_2 ^2 G_{X2} ^2 + v_S ^2 g_X ^2 q_X ^2 \end{pmatrix} \,.
\label{mixinzx}
\end{equation}
The above expression, Eq.\ \eqref{mixinzx}, representing the mixing between the $Z^0 _\mu$ and $X_\mu$ bosons, is given as function of arbitrary $U(1)_{X}$ charges of doublet (or singlet) scalars. It is important to notice that, when $Q_{X1}=Q_{X2}$ and there is no singlet contribution, the determinant of the matrix Eq.\ \eqref{mixinzx} is zero.

The matrix in Eq.\ \eqref{mixinzx} is diagonalized through a rotation $O(\xi)$
\begin{equation}
\label{rotacao_zz_fisicos}
\begin{pmatrix} Z_\mu \\ Z ' _\mu \end{pmatrix} = \begin{pmatrix} \cos \xi & - \sin \xi \\ \sin \xi & \cos \xi \end{pmatrix} \begin{pmatrix} Z^0 _\mu \\ X_\mu \end{pmatrix}
\end{equation}
and its eigenvalues are
\begin{equation}
\begin{split}
\label{autovalores_matriz_zz}
m_{Z} ^2 &= \frac{1}{2} \left[ m_{Z ^0} ^2 + m_X ^2 - \sqrt{ \left( m_{Z ^0} ^2 - m_X^2 \right) ^2 + 4 \left( \Delta ^2 \right) ^2} \right], \\
m_{Z '} ^2 &= \frac{1}{2} \left[ m_{Z ^0} ^2 + m_X ^2 + \sqrt{ \left( m_{Z ^0} ^2 - m_X^2 \right) ^2 + 4 \left( \Delta ^2 \right) ^2} \right].
\end{split}
\end{equation}
The $\xi$ angle is given by
\begin{equation}
\tan \xi = \frac{ \Delta ^2}{m^2 _{Z^0} - m^2 _{X}} .
\label{Eq:xi}
\end{equation}
Since this mixing angle it supposed to be small, as  $m_{Z^\prime}^2 \gg m_Z^2$, we can use $\tan \xi \sim \sin \xi$ with
\begin{equation}
\sin \xi \simeq \frac{ G_{X1} v_1^2 + G_{X2} v_2^2}{m^2_{Z^\prime}} .
\label{Eq:xinew}
\end{equation}
We can expand this equation further to find a more useful expression. Substituting the expressions for $G_{Xi}$ and factoring out the $m_Z$ mass, we finally get

\begin{equation}
\sin \xi \simeq \frac{m_Z^2}{ m^2_{Z^{\prime}}}\left(\frac{g_X}{g_Z}(Q_{X1}\cos^2 \beta + Q_{X2}\sin^2 \beta) +\epsilon \tan\theta_W \right).\\
\label{eqsinxi}
\end{equation}

\section*{$Z$ Properties and LEP Bounds}\label{AppendixZ}

\begin{figure*}[h!]
\includegraphics[width=\columnwidth]{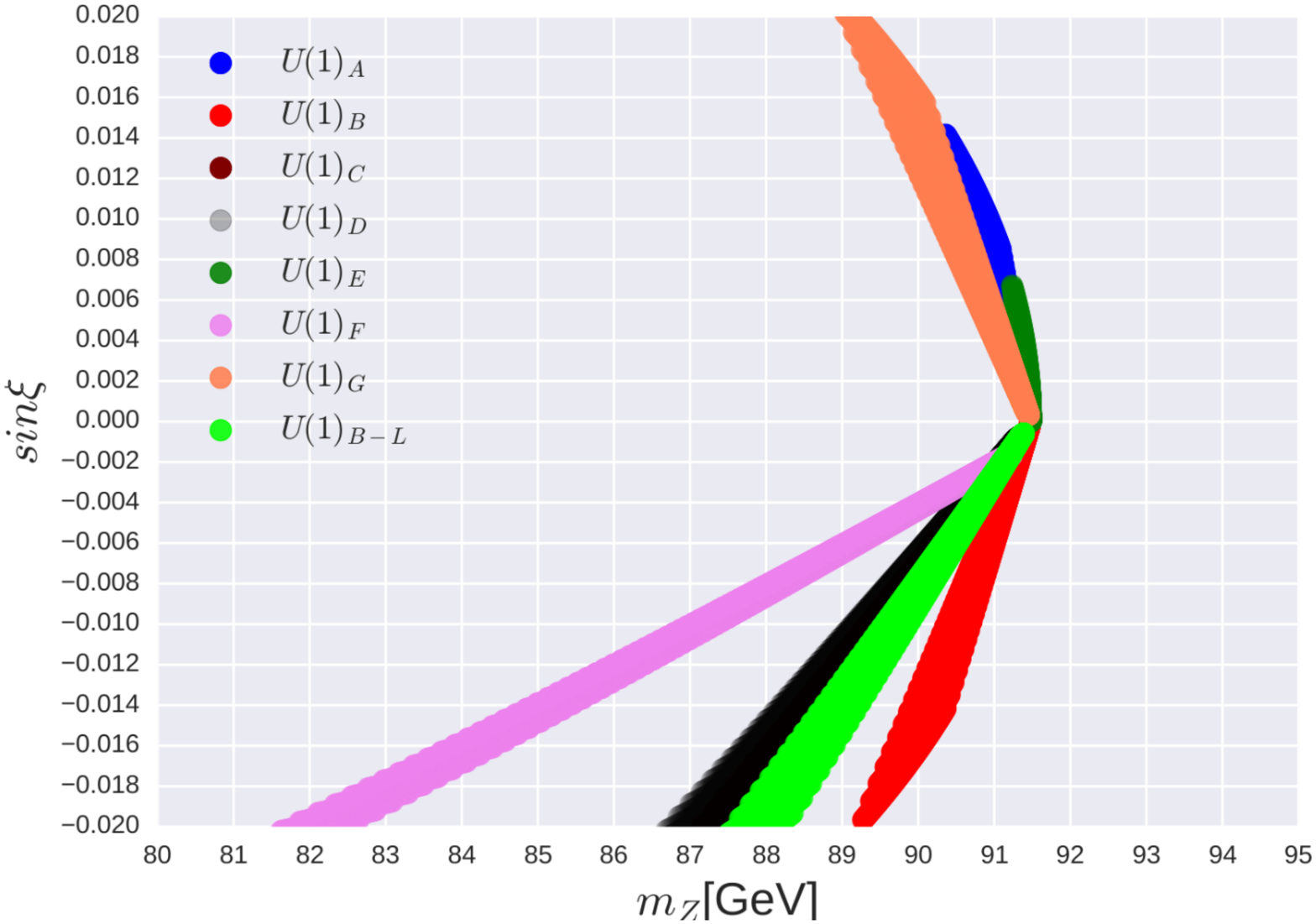}
\includegraphics[width=1.05\columnwidth]{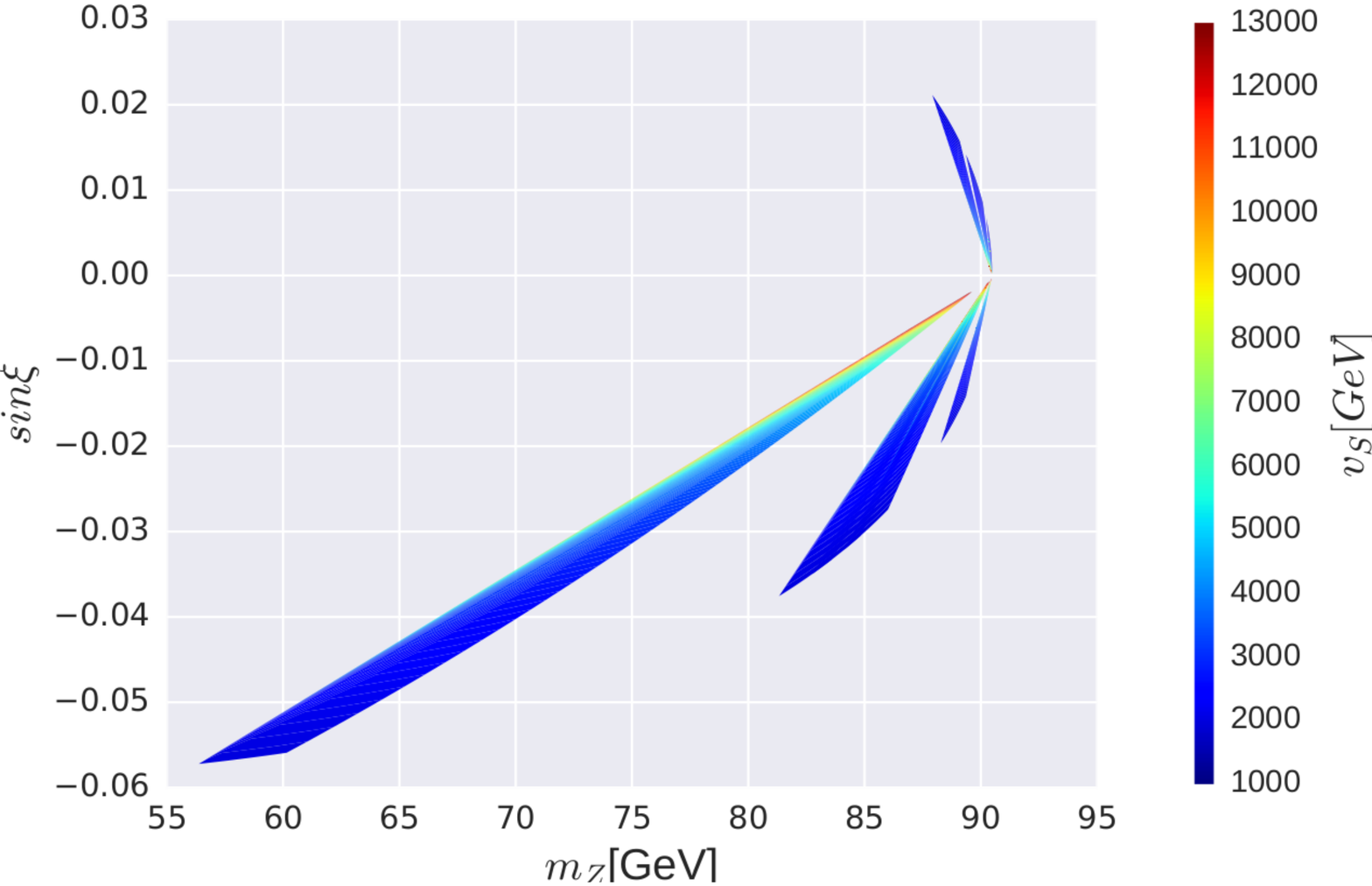}
\caption{(Left) SM $Z$ mass behavior with $\sin\xi$ for all the models considered here. (Right) Same with the addition of a color map that represents how $v_S$ varies for all  $U(1)_X$ models considered.}
\label{fig:LEP_2}
\end{figure*}

All the $U(1)_X$ configurations demand high $v_S \gtrsim 5$~TeV values in order to be in agreement with the SM $Z$ mass measurements of $m_Z = 91.1876\pm 0.0021$. This has been taken into account and the LEP limits showed here encompass this constraint. We compiled them in table \ref{tabLEP} to ease the reading. In this procedure we adopted $\epsilon=10^{-3}$ to be consistent with EW precision data \cite{Carena:2004xs} and $\tan \beta = v_2/v_1 = 10$ in agreement with the latest experimental constraints on the 2HDM \cite{Bernon:2015wef}. In the {left panel} of figure \ref{fig:LEP_2} we showed how the $Z$ mass changes depending on the mixing angle adopted for every single $U(1)_X$ model. This illustrates that, indeed, only small mixing angles of the order of $10^{-3}$ are allowed by LEP data. In the {right panel} of the same figure, for completeness, we exhibit a heat map to indicate which value of $v_S$ is needed to reproduce a small mixing. Using LEP precision data and the theoretical predictions discussed above, we can estimate that $v_S$ should be larger than $\sim 10$~TeV for all models investigated here.\\

\begin{figure*}[t!]
\includegraphics[width=\columnwidth]{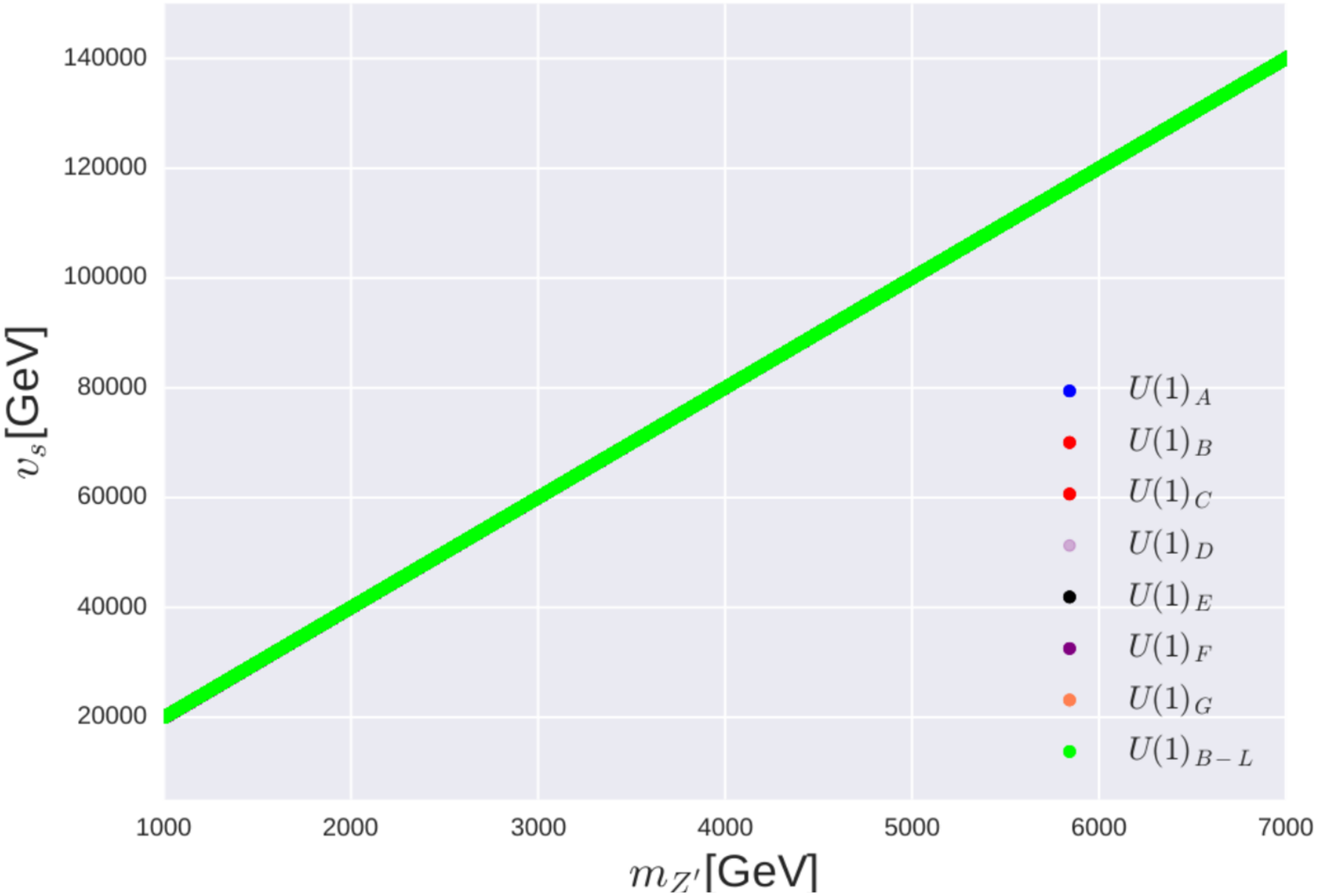}
\includegraphics[width=\columnwidth]{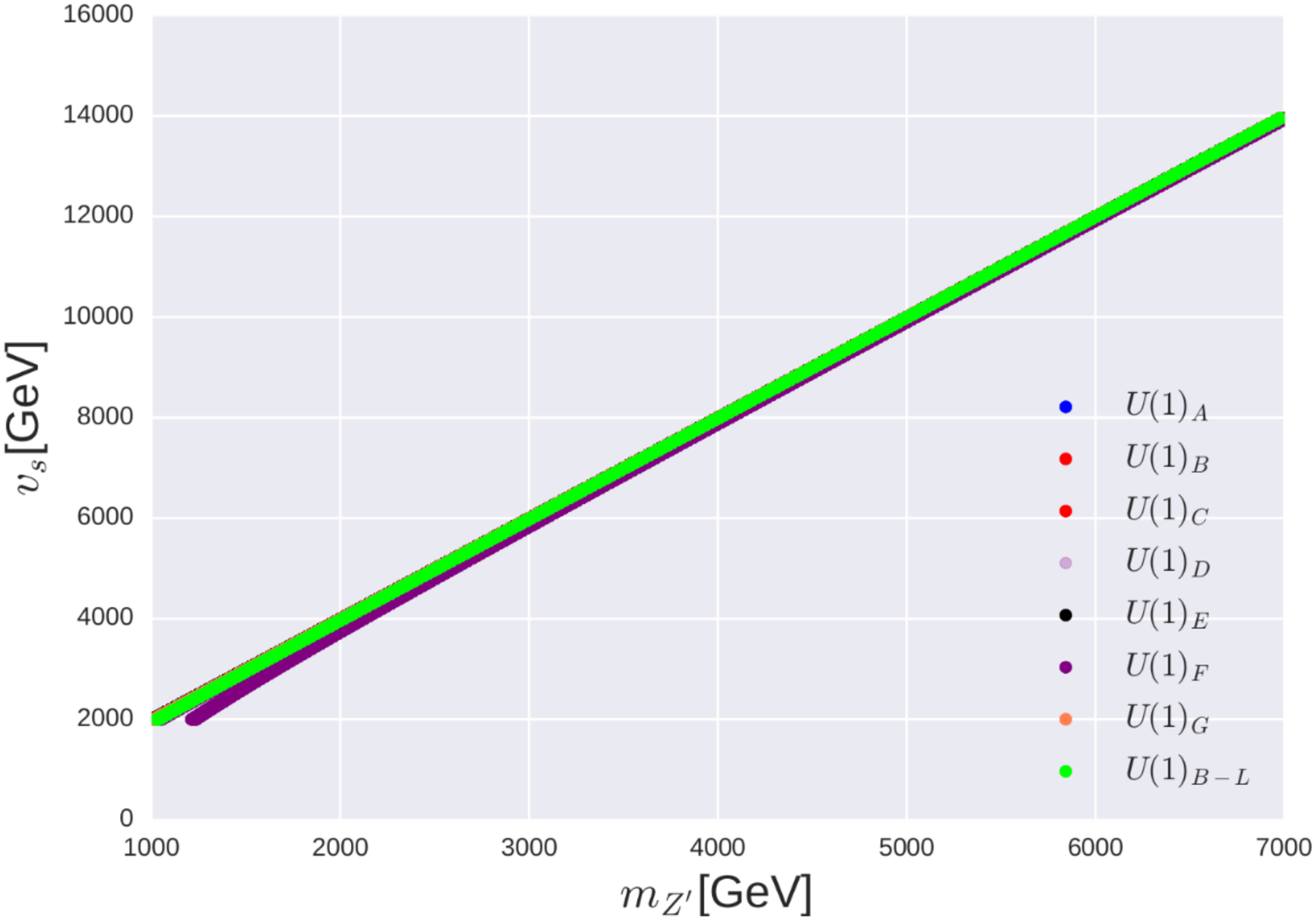}
\caption{We show here the $m_{Z^\prime}$ dependence on $v_S$ for $g_X=0.1$ (left) and $g_X=1$ (right), assuming $\epsilon=10^{-3}$ and $\tan\beta =10$, for all  $U(1)_X$ models considered.}
\label{fig:mzp}
\end{figure*}

\section*{$Z^\prime$ Width}\label{AppendixZprime}

\begin{figure*}[t!]
\includegraphics[width=\columnwidth]{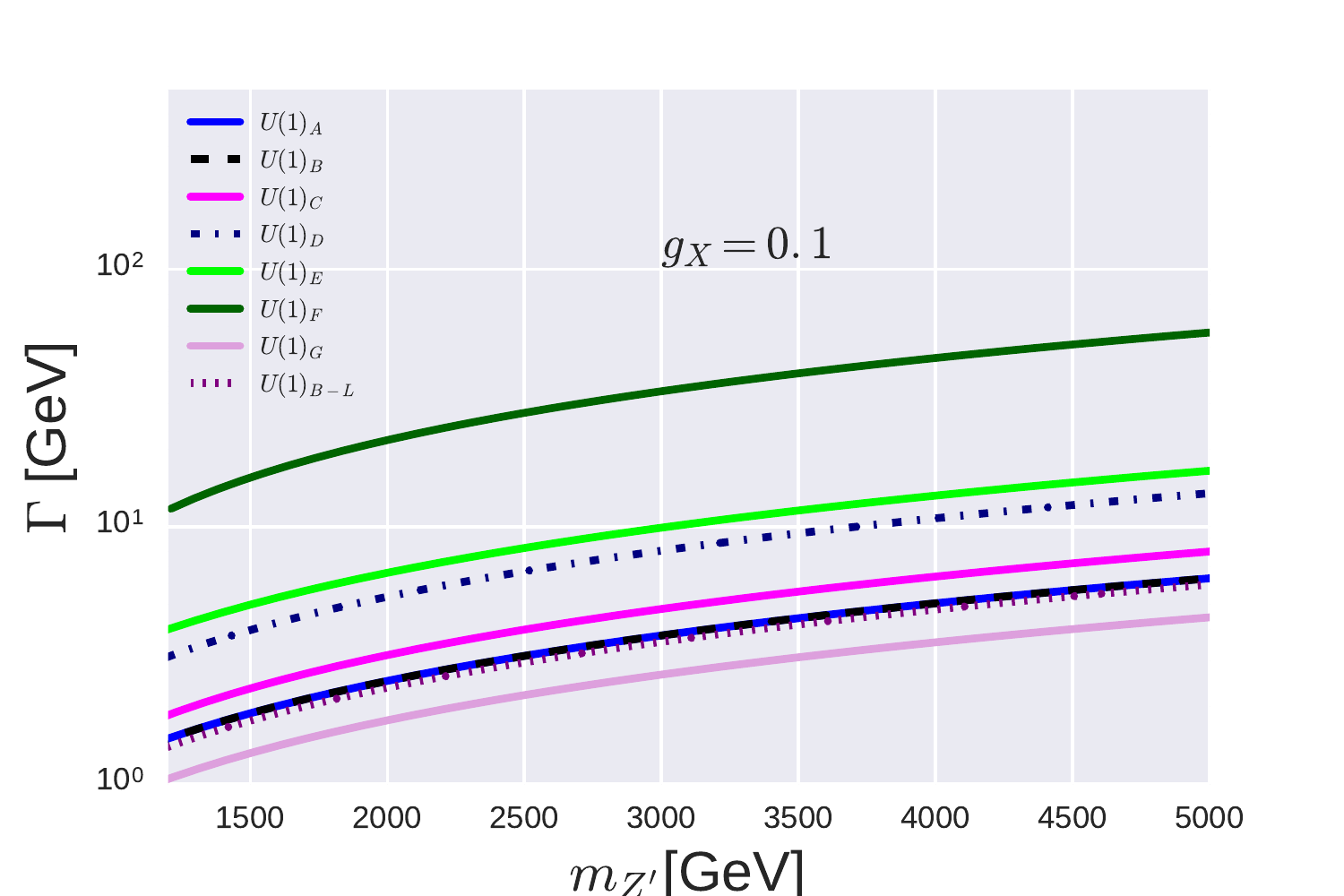}
\includegraphics[width=\columnwidth]{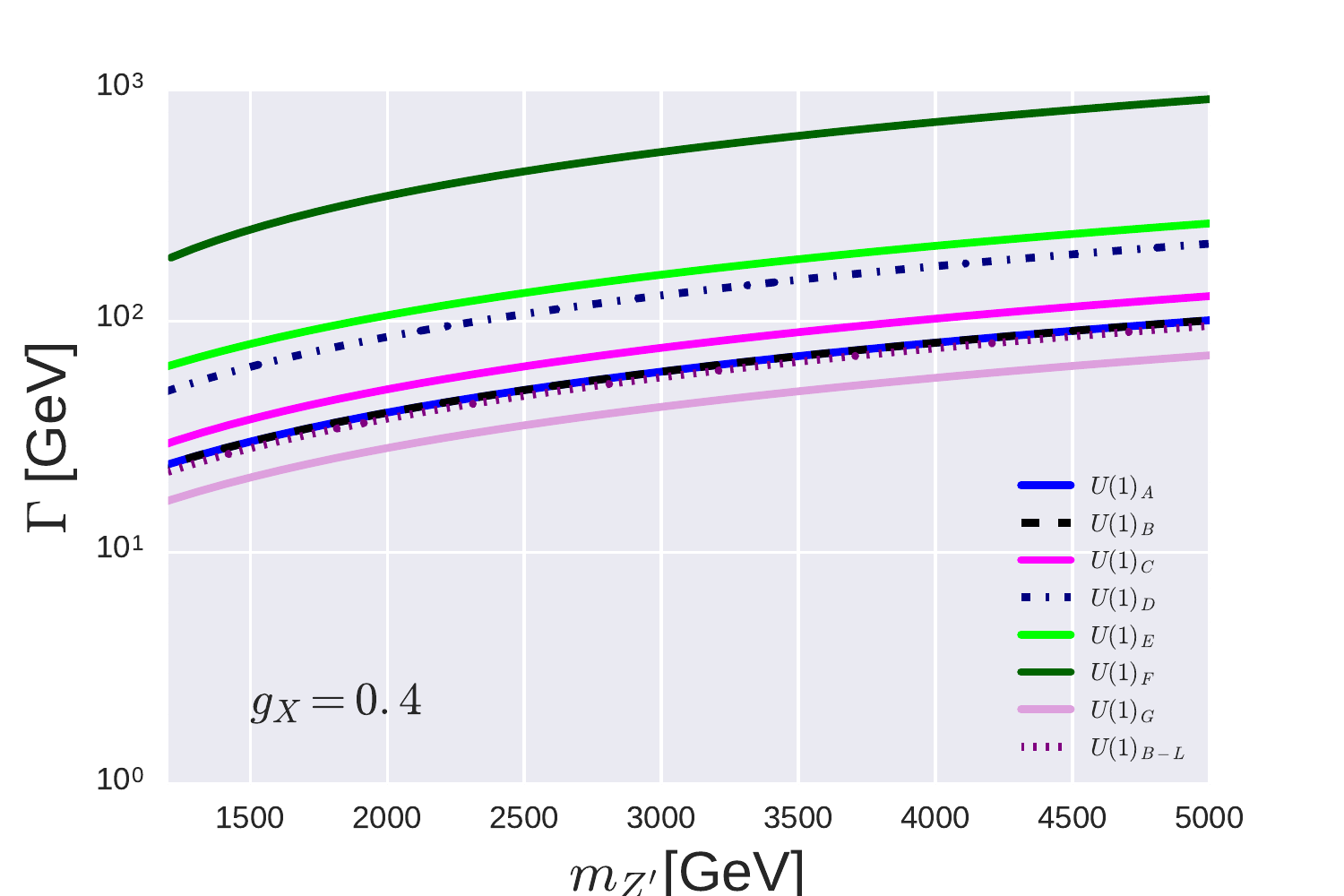}
\includegraphics[width=\columnwidth]{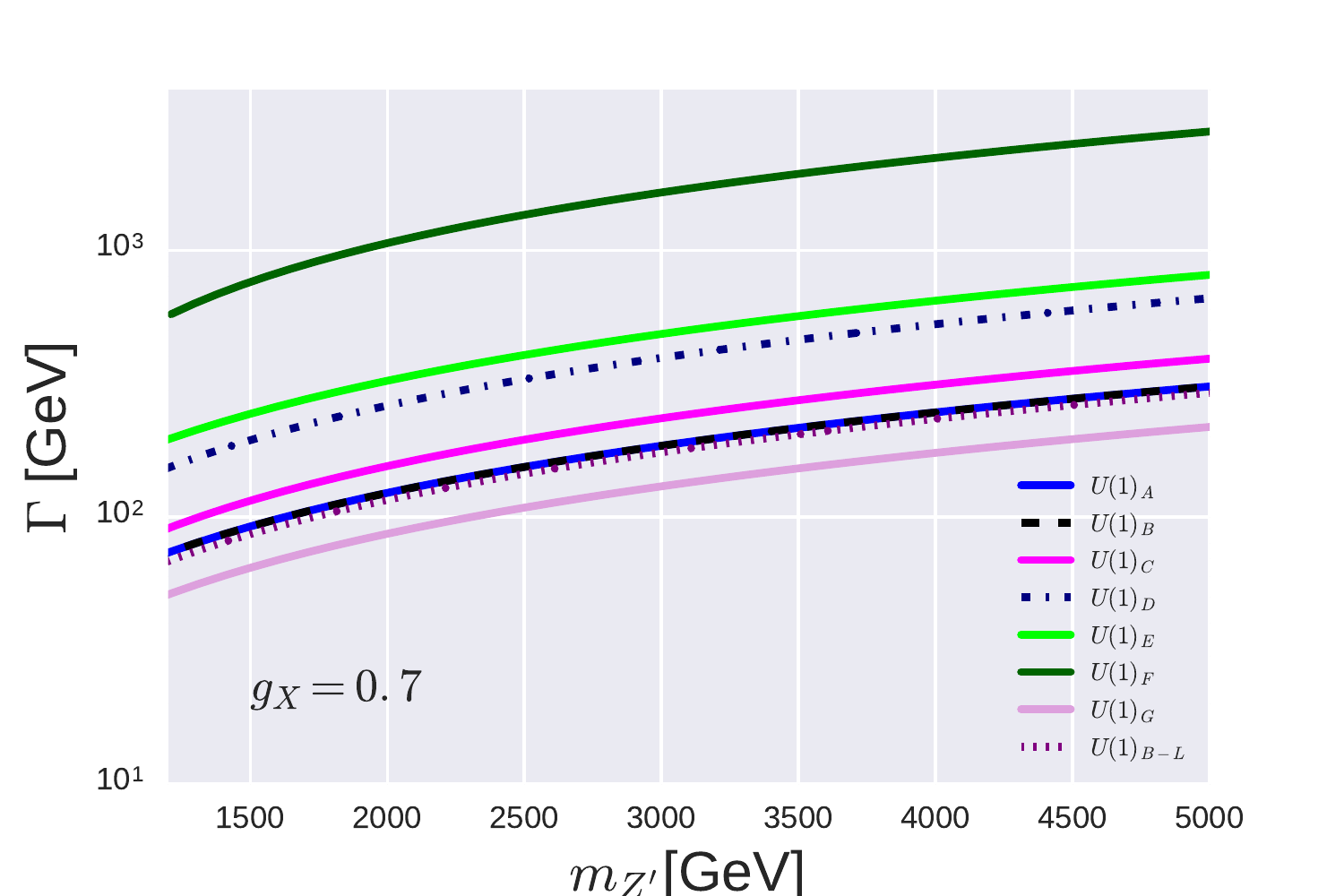}
\includegraphics[width=\columnwidth]{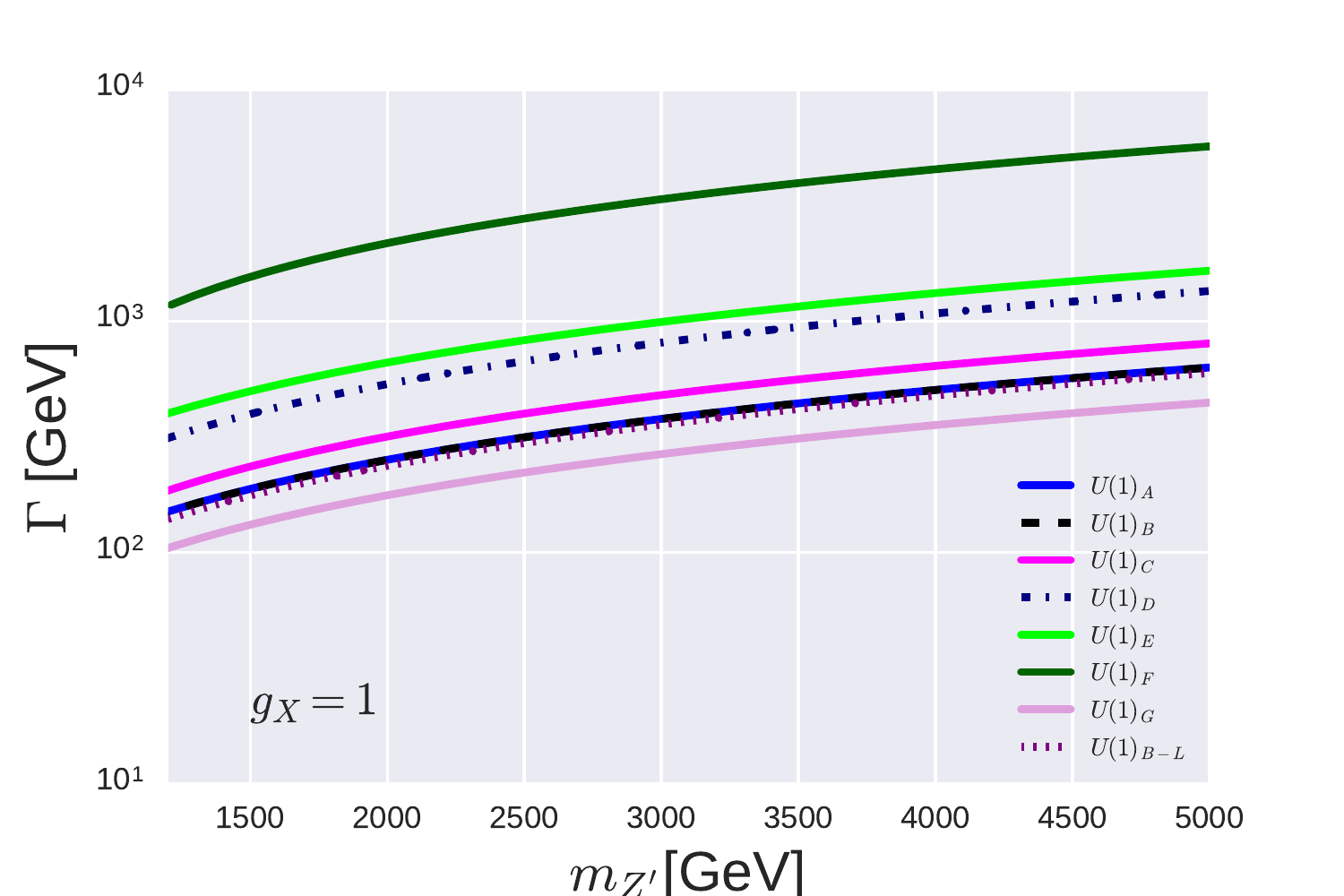}
\caption{We show here the $Z^\prime$ width as function of its mass for several values of $g_X$ for all  $U(1)_X$ models considered in this paper.}
\label{fig:width}
\end{figure*}

In figure \ref{fig:mzp} we show how the $Z^\prime$ mass scales with $v_S$ for $g_X=0.1$ and 
$1$. It is clear that, irrespectively of  the details of the model, as soon as $v_S^2 \gg 246^2$~GeV, the $Z^\prime$ mass is dominated by $v_S$ and  scales linearly with it.
In figure \ref{fig:width} we show the $Z^\prime$ width as a function of its mass for $g_X=0.1, 0.4, 0.7$ and $1$, where one can easily see that, for increasing $g_X$, the $Z^\prime$ width eventually becomes too large relative to the mass so that one cannot use the NWA to derive LHC bounds on the $Z^\prime$ properties. 

\bibliography{literature}

\end{document}